\documentclass[11pt, draftclsnofoot, peerreview, onecolumn]{IEEEtran}

\IEEEoverridecommandlockouts

\usepackage{lineno}
\usepackage{amssymb}
\usepackage{amsmath}
\usepackage{amsthm}
\usepackage{epsfig}
\usepackage{graphicx}
\usepackage{graphics}
\usepackage{float}
\usepackage{subfigure}
\usepackage{multirow}
\usepackage{color}
\usepackage{fullpage}
\usepackage[normalem]{ulem}
\usepackage{makeidx}
\usepackage{xspace}
\usepackage{wrapfig}
\usepackage{lipsum}
\usepackage{mathtools}
\usepackage{cuted}
\usepackage{cite}
\usepackage{amsfonts}
\usepackage{algorithmic}
\usepackage{textcomp}
\newcommand{\diag}{\mathop{\mathrm{diag}}}

\makeindex
\theoremstyle{plain}
\newtheorem{theorem}{{ Theorem}}

\theoremstyle{definition}
\newtheorem{Definition}{Definition}

\newtheorem{remark}{Remark}
\newtheorem{lemma}{Lemma}

\makeatletter
\renewcommand\@endtheorem{\vvv@endmarker\endtrivlist\@endpefalse}
\newcommand\vvv@endmarker{%
  {\unskip\nobreak\hfil\penalty50
  \hskip2em\vadjust{}\nobreak\hfil\openbox
  \parfillskip=0pt \finalhyphendemerits=0 \par
  \penalty 10000 \parskip=0pt\noindent}\ignorespaces}
\makeatother
\newtheorem{example}{Example Part}
\definecolor{darkred}{rgb}{1, 0.1, 0.3}
\definecolor{darkblue}{rgb}{0.1, 0.1, 1}
\definecolor{darkgreen}{rgb}{0,0.6,0.5}
\def\BibTeX{{\rm B\kern-.05em{\sc i\kern-.025em b}\kern-.08em
    T\kern-.1667em\lower.7ex\hbox{E}\kern-.125emX}}

\newcommand\bovermat[2]{%
  \makebox[0pt][l]{$\smash{\overbrace{\phantom{%
    \begin{matrix}#2\end{matrix}}}^{\text{#1}}}$}#2}
\def \A {\mathcal{A}}
\def \S {\mathcal{S}}
\def \Z {\mathcal{Z}}
\def \D {\mathcal{D}}

\def \T {\mathcal{T}}

\def \I {\mathcal{I}}
\def \J {\mathcal{J}}
\def \F {\mathbb{F}}

\allowdisplaybreaks

\begin{document}
\cleardoublepage
\setcounter{page}{1}
\title{The Capacity Region of \\ Distributed Multi-User Secret Sharing}

\author{\IEEEauthorblockN{
%Author1, Author2, Author3, Author4
Ali Khalesi, Mahtab Mirmohseni, and Mohammad Ali Maddah-Ali}\\
\IEEEauthorblockA{Sharif University of Technology}}

\maketitle

\begin{abstract}
In this paper, we study the problem of distributed multi-user secret sharing, including a trusted master node, $N\in \mathbb{N}$ storage nodes, and $K$ users, where each user has access to the contents of a subset of storage nodes.
Each user has an independent secret message with certain rate, defined as the size of the message normalized by the size of a storage node.  Having access to the secret messages, the trusted master node places encoded shares in the storage nodes, such that (i) each user can recover its own message from the content of the storage nodes that it has access to, (ii) each user cannot gain any information about the message of any other user.  We characterize the capacity region of the distributed multi-user secret sharing, defined as the set of all achievable rate tuples, subject to the correctness and privacy constraints. In the achievable scheme, for each user, the master node forms a polynomial with the degree equal to the number of its accessible storage nodes minus one, where the value of this polynomial at certain points are stored as the encoded shares.  The message of that user is embedded in some of the coefficients of the polynomial. The remaining coefficients are determined such that the content of each storage node serves as the encoded shares for all users that have access to that storage node.
\end{abstract}


\section{Introduction}

The new information era in which we are living on necessities the use of distributed systems for the storage and processing tasks of large amount of data, realized by cloud computing and storage services. The distributed nature of these services along with supporting multiple users makes the security and privacy concerns one of the most important challenges we face.
The users of cloud services expect their data to be secure and private guaranteeing that only the intended user is able to recover its data, while the service providers seek the efficient algorithms reducing the required memory, communication overhead, delay, and processing costs.

\begin{figure}[t]
\centering
\includegraphics[scale=0.7]{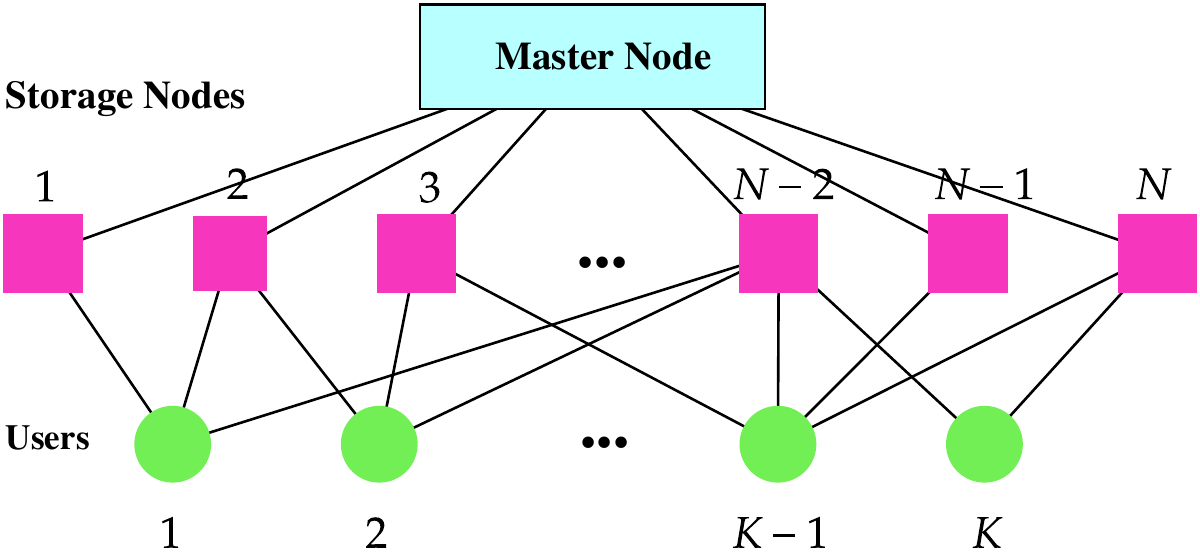}
\caption{A schematic of a simple \textbf{DMUSS} system, consisting of a master nodes, some storage nodes and users.}\label{fig:0}
\end{figure}
A classic solution for this challenge in the context of distributed storages may make use the secret sharing protocols.
The secret sharing scheme, introduced by Shamir \cite{r4} and Blakely \cite{blakley1979safeguarding} in 1979, is a prominent building block for many cryptographic applications, such as threshold cryptography \cite{desmedt1991shared,desmedt1989threshold,shoup2000practical}, secure multiparty computation \cite{goldwasser1988completeness,chaum1988multiparty, cramer2000general,Akbari,lagrange},  secure private information retrieval \cite{yang2018private2},   secure matrix processing \cite{chang2018capacity, d2020gasp,aliasgari2020private}, Byzantine agreement \cite{rabin1983randomized}, access control \cite{naor1998access} and  attribute-based encryption \cite{goyal2006attribute,waters2011ciphertext}.

Using secret sharing schemes in securing the distributed cloud services allows to store and communicate the secret shares of each user's data. However, the cloud services are supposed to serve multiple users in a shared platform and thus the security and privacy of user's data must be protected in a distributed multi-user system. To adopt secret sharing schemes to this scenario, secret shares of users must be stored in the distributed servers and privately communicated to them, who might have access to some of the distributed servers with different access structures.
Thus, we encounter a resource sharing problem, in which the storage must be shared by the different users. A naive solution is to split the storage between the users and to use separate secret sharing for each user. However, this incurs a very large overhead to the system in a multi-user setup as it needs many shares generated per user's secret message.
An interesting question here is that whether it is possible to jointly allocate the resources to the users by jointly producing their secret shares and gain order-wise improvement over storage splitting. This problem has been first studied by Soleymani and Mahdavifar \cite{soleymani2020distributed}; where their focus is on the cases where the access structure has a regularized form  and the users secret messages have equal size.

The distributed multi-user secret sharing (DMUSS) problem consists of a trusted master node/dealer, a set of $N\in \mathbb{N}$ equal-size storage nodes and a set of $K\in \mathbb{N}$ users (see Fig.~\ref{fig:0}). The master node is directly connected to all storage nodes through some error-free, reliable and secure links. Furthermore, each user has access to a specific subset of storage nodes, called the access set of that user,  and is able to read their stored data without error. Each user has its own secret message of certain size.
The master node has access to all secret messages. The master node aims to store some encoded shares in the storage nodes such (i) each user can recover its own message  correctly from the contents of the storage nodes that it has access to, (ii) each user cannot gain any information about the message of any other user.

For more intuition, consider a simple setup illustrated in Fig.~\ref{fig:0}.
\begin{figure}
\centering
\includegraphics[scale=0.7]{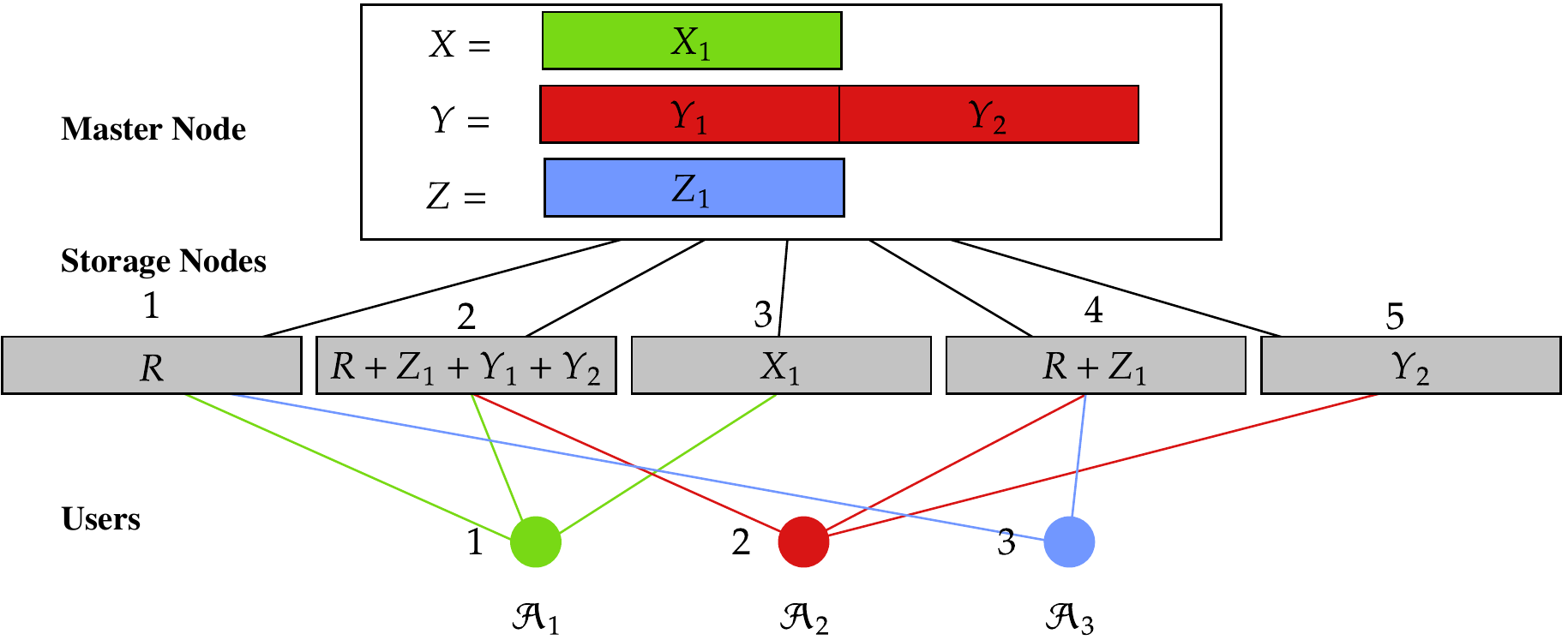}
\caption{A motivating example consisting of a master node, 5 storage nodes and 3 users. The master node stores 3 secret messages, $X,Y,Z$. Each user desires its secret message correctly and privately.}\label{fig:00}
\end{figure}
Users 1, 2, and 3 wish to retrieve $X \in \mathbf{GF}(q)$, $Y \in (\mathbf{GF}(q))^{2}$ and $Z \in \mathbf{GF}(q)$, respectively, where $\mathbf{GF}(q)$ denotes a finite field of size $q$.  User 1 has access to storage nodes $\{1,2,3\}$, User 2 has access to storage nodes $\{2,4,5\}$, and User 3 has access to storage nodes $\{1,4\}$.
To provide privacy, the master node chooses a uniformly and independently distributed random value $R \in \mathbf{GF}(q)$. To decode the secret messages, User 1 can retrieve its message directly from storage node 3; User 2 recovers $Y_2$ from storage node 5, and $Y_1$ by subtracting the contents of storage nodes 4 and 5 from the content of storage node 2; and User 3 subtracts the content of storage node 1 from the content of storage node 4. For privacy, note that no other user has access to storage node 3, therefore message $X$ will be private. With the same reason for User 2, $Y_2$ is private and also User $1$ is not able to learn anything about $Y_1$,  since it does not know $Z_1, Y_2$. $Z_1$ is also concealed from Users 1 and 2, since no user has access to both storage nodes 1 and 4 and $R$. Thus the privacy and correctness are preserved.

In this paper, we investigate the fundamental limits of the distributed multi-user secret sharing systems with arbitrary access structure. In this set-up, users can have different message sizes. Since the storage nodes are shared among different users, as we increase the message size of one user, the size of the messages of other users may be reduced. Let us define the rate of each user as the size of its message normalized by the size of a storage node. Then the question of interest is to know which rate tuples  are achievable. In this paper, we define the notation of capacity region for the distributed multi-user secret sharing, as the set of all achievable rate tuples, subject to the correctness and privacy constraints. We characterize the capacity region, by developing an achievable scheme and proposing a matching converse.  The main idea of the  the achievable scheme is as follows:  For each individual user, the master node  forms a polynomial with the degree equal to the size of its access set minus one. The value of this polynomial at certain points are stored as the encoded shares.  The message of that user is partitioned and used as some of the coefficients of the polynomial. The remaining coefficients are determined such that the content of each storage node serves as the encoded shares for all users that have access to that storage node, forming a set of linear equations. We prove that if the rate vector satisfies certain conditions, this set of equations has a solution. Moreover, each user has enough ambiguity about the message of any other user and learns nothing about it.

This paper is motivated by the result of \cite{r1}. However, in comparison, the proposed scheme here  benefits  from the several desirable features, (i) it allows different message sizes for the users, (ii) it works with arbitrary access structure, (iii) it is capacity achieving.

\noindent{\bf Notations:}
For $n\in\mathbb{N}$, define $[n]$ as the set $\{1,2,\hdots,n\}$, and for $n_1,n_2\in\mathbb{Z},n_1\leq n_2$, define $[n_1:n_2]$ as the set $\{n_1,n_1+1,\hdots,n_2\}$. For a set $\mathcal{I}=\{i_1,i_2,\hdots,i_n\}$, $A_{\mathcal{I}}$ represents $\{A_{i_1},A_{i_2},\hdots,A_{i_n}\}$. For two sets $\mathcal{A}_1$ and  $\mathcal{A}_2$, $\mathcal{A}_1\backslash\mathcal{A}_2$ is the set of elements that are in $\mathcal{A}_1$ but not in $\mathcal{A}_2$. For matrices $\mathbf{A}$ and $\mathbf{B}$, $[\mathbf{A},\mathbf{B}]$ indicates the concatenation of two matrices. $o(M)$ represents a function that approaches zero as $M\rightarrow\infty$. $\mathbf{I}_K$ represents $K\times K$ identity matrix. If $\mathbf{C}$ is a matrix and $\I,\J \in \mathbb{N}$, then $\mathbf{C}(\I,\J)$ is the sub-matrix resulted from the row indices in $\I$ and columns in $\J$. We denote the finite field $\mathbf{GF}{(q^M)}$ as $\mathbb{F}$.
For a vector $\mathbf{w}=[w_{0},w_{1},\hdots,w_{R-1}]^\intercal \in \F^{R\times 1}, R \in \mathbb{N}$, we define the polynomial   $g_{\mathbf{w}}(x)$ as
\begin{equation}
    g_{\mathbf{w}}(x)\triangleq\sum^{R-1}_{r=0}w_{r}x^r \label{4-2-1}.
\end{equation}
Also, for $\mathbf{w}_1 \in \F^{R_1}, R_1 \in \mathbb{N}$ and  some $\mathbf{w}_2 \in \F^{R_2}, R_2 \in \mathbb{N}$, we define the polynomial $g_{\mathbf{w}_1,\mathbf{w}_2}(x)$ as
\begin{equation}
    g_{\mathbf{w}_1,\mathbf{w}_2}(x)\triangleq g_{\mathbf{w}_k}(x)+x^{R_1} g_{\mathbf{w}_2}(x).\label{4-2-2}
\end{equation}

\section{System Model}
\label{sec: System Model}

\begin{figure}[t]
\centering
\includegraphics[scale=0.8]{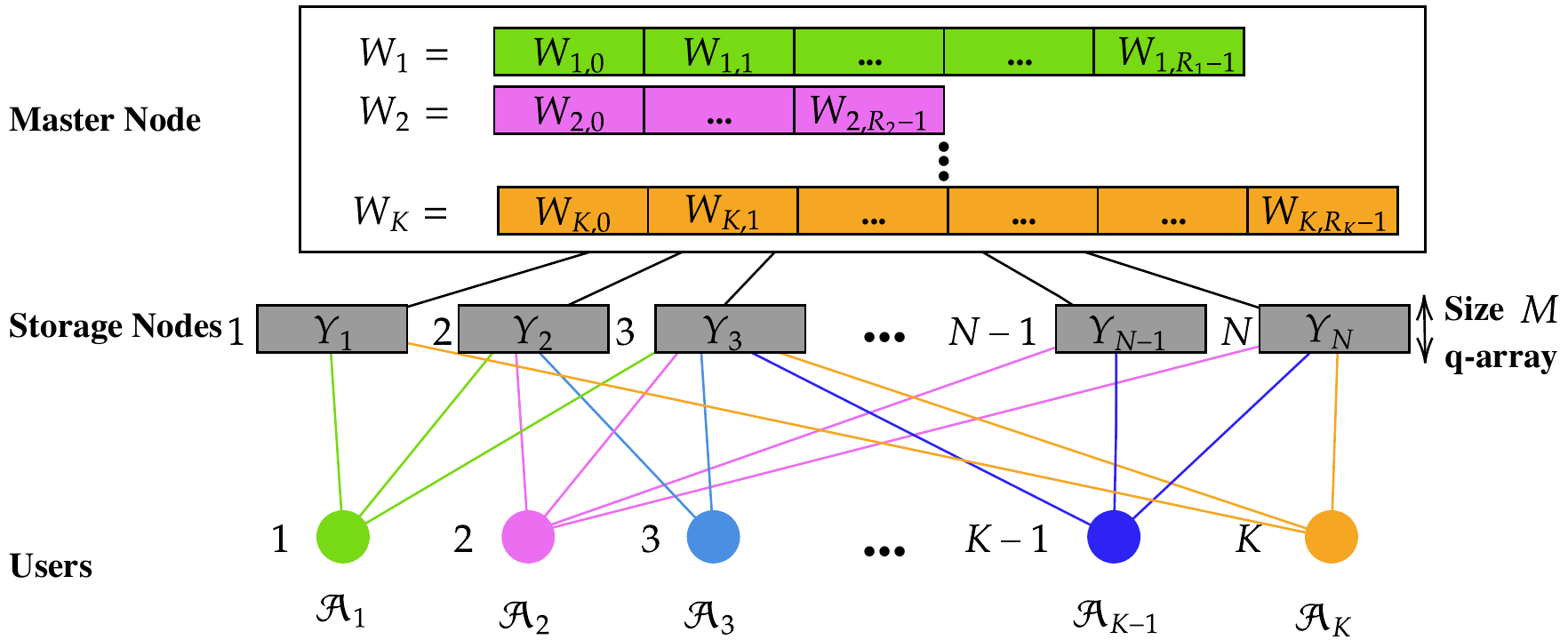}
\caption{The distributed multi-user secret sharing (DMUSS) system: A master node containing $K$ secret messages $W_k,\: k \in [K]$ of size $r_k , k\in [K]$, which is connected through error-free links to $N$ storage nodes each with a storage of size $M$. The $k$-th user is connected to the storage nodes in $\A_k$. The goal is to design the placement phase and the retrieval phase such that the correctness and privacy conditions be satisfied.}\label{fig:1}
\end{figure}

The DMUSS problem consists of a master node, $N\in\mathbb{N}$ storage nodes and $K\in\mathbb{N}$ users. The $k$-th user (for $k\in[K]$) has its own secret message $W_k$. All secret messages, i.e., $W_1,W_2\hdots,W_K$, are stored in the master node. The master node is connected to all storage nodes, each with storage size $M\in\mathbb{N}$. The $k$-th user is connected through some error-free links to a set of storage nodes $\mathcal{A}_k\subseteq[N],\forall k\in[K]$, as shown in Fig.~\ref{fig:1}. The users intend to retrieve their own secret messages correctly and privately from their connected storage nodes.
\begin{Definition}\label{Def1}
Denoting the length of $W_k$ with $r_k$, we define the $k$-th user secret message rate as:
\begin{equation}
R_k\triangleq\frac{r_k}{M}.\label{2-1}
\end{equation}
\end{Definition}

\textbf{Placement and retrieval phases}:
The system works in two phases. In the placement phase, the master node places the encoded shares of size $M$ $q$-array bits, denoted as $Y_n$, in the $n$-th storage node, for $n\in[N]$. In the retrieval phase, each user $k$ must be able to retrieve the secret message $W_k$ \emph{correctly} and \emph{privately} through its unique access set $\mathcal{A}_k$, imposed by the topology of the network.

Let $W_1,W_2\hdots,W_K$ be $K$ mutually independent secret messages, where each $W_k,\: k \in [K]$ is uniformly distributed over $\mathbb{F}^{\lceil R_k \rceil\times 1}$ where $k\in[K],R_k\in\mathbb{R^+}\cup\{0\},M\in\mathbb{N}$. So,
\begin{align}
  \textrm{[independence]} \:\:\:\:  H(W_1,W_2,\hdots,W_K)&= \sum^K_{i=1}H(W_i),
    \label{2-2}\\
     \textrm{[length]}\:\:\:\: \:\:\:\:\:\:\:\:\:\: H(W_k)=R_kM&=r_k\:\:\:\:\:\forall{k\in[K]}.
    \label{2-3}
\end{align}
In the placement phase, the master node  encodes the stored secret messages with an encoding function $\phi^{M}$, that maps the secret messages $W_1,\hdots,W_K$ to $Y_1,\hdots,Y_N$, as
\begin{equation}
\phi^{M}:\prod^{K}_{i=1} \mathbb{F}^{R_i} \xrightarrow{}\prod^{N}_{i=1} \mathbb{F}.
    \label{2-4}
\end{equation}
$Y_n,\: n \in [N]$ represents the content of the $n$-th storage node after the placement phase.
In the retrieval phase, each user $k\in[K]$ applies the decoding function $\theta^{M}_{k}$ to retrieve $W_k$ from its accessible stored data, $Y_{\mathcal{A}_k}$, as
\begin{equation}
\theta^{M}_{k}:\prod^{{|\A_k|}}_{i=1} \mathbb{F} \xrightarrow{}\mathbb{F}^{R_i},
    \label{2-5}
\end{equation}
which maps the contents of the $k$-th user's accessible storage nodes, $Y_{\mathcal{A}_{k}}$, to a decoded secret message $\hat{W}_k$.
The encoder and decoder functions \eqref{2-4} and \eqref{2-5} must satisfy correctness and privacy conditions, stated below.

\textbf{Correctness and privacy conditions}:
The correctness condition is defined to ensure the ability of User $k$ to reconstruct the secret message $W_k$ with arbitrary small error, as
\begin{equation}
  \lim_{M\xrightarrow{} \infty} \mathbb{P}_e=0, \label{2-6}
\end{equation}
where $P_e=\mathbb{P}(W_k\neq\hat{W}_k)$ represents the probability of decoding error. By Fano's inequality, we obtain
\begin{equation}
\frac{1}{M}H(W_k|\hat{W}_k)= o(M).\label{2-6-1}
\end{equation}

The privacy condition means that each user $k\in[K]$ must not learn any information about any other secret message $W_{\tilde{k}},\: \tilde{k}\in[K]\backslash \{{k}\},$ observing its access set $\A_k$. For any $k,\tilde{k}\in[K],$ $k\neq \tilde{k}$, we must have,
    \begin{equation}
H(W_{\tilde{k}})=H(W_{\tilde{k}}|Y_{\mathcal{A}_k}).   \label{2-7}
\end{equation}

\begin{Definition}\label{Def:capacity}
The rate tuple $(R_1,R_2,\hdots,R_K)$ is achievable if there exists a sequence of \textbf{DMUSS} schemes (consisting of the encoder and decoder function that satisfy the correctness condition in \eqref{2-6} and privacy condition in \eqref{2-7}) with the rate of $k$-th secret message equal to $R_k$. The capacity region $\mathcal{C}$ of DMUSS is defined as the closure of the set of all achievable rate tuples.
\end{Definition}

\section{Main Results: The Capacity Region of Distributed Multi-User Secret Sharing}
\label{sec:main results}
We characterize the capacity region of DMUSS in the following theorem.
\begin{theorem}\label{Theorem 1}
The capacity region of DMUSS is the convex hull of all regions with the rate tuple $(R_1,\hdots,R_K)$ satisfying:
\begin{align}
      R_k & \leq \min_{k  \neq \tilde{k}}|\A_k\backslash \A_{\tilde{k}}|,\:\: \forall{k,\tilde{k}\in[K]},\label{3-1}\\
  \textrm{[cutset condition]}\:\:\:\:  \sum^{}_{i\in \S}R_i & \leq|\cup_{i\in \S} \A_i|,\:\:\:\: \S\subseteq[K].
   \label{3-2}
\end{align}
\end{theorem}
The achievability and converse proofs are provided in Section \ref{sec: Achievabilty 1} and Section \ref{sec: Converse 1}, respectively.
\begin{remark}
The following intuitive remarks about Theorem~\ref{Theorem 1} can be concluded.
\begin{enumerate}
\item The achievable scheme depends on a polynomial for each user, defined to realize the encoding function, whose degree is determined by the size of the corresponding access set. Some of its coefficients are chosen according to the privacy condition and a system of linear equations. The rest are equal to the corresponding secret message or are determined by a system of linear equations requirements. The system of linear equations takes its shape according to these polynomials and the access set of each user. The feasibility of the encoding function requires the existence of a unique feasible solution for this system of linear equations, which is proved when \eqref{3-1} and \eqref{3-2} holds.
\item As can be seen from \eqref{3-1}, if there exists a User $\tilde{k}$ which its accessible set has a large number of common elements with User $k$, the rate  $R_k$ would be small. Intuitively speaking, the more denser the access set be, the less secret message rate's average become.
\item The result of Theorem \ref{Theorem 1} includes the result of \cite{r1} as a special case. The achievable scheme proposed in \cite{r1} considers DMUSS problem when the access sets have a specific structure and also the focus was on the feasibility of rate-tuple $(R_1,R_2,\hdots, R_k)=(1,1,\hdots,1)$. In particular, the achievable access sets  preserve the condition $|\A_k\backslash \A_{\tilde{k}}| =1,\: \forall{k,\tilde{k}}\in[K]$  and also are assigned by an algorithm to have a regularized structure. However, our results are valid for any arbitrary access structure. In addition, we investigate the feasibility of any rate-tuple $(R_1,R_2,\hdots, R_k)$.
\end{enumerate}
\end{remark}

\section{The Achievability Proof}
\label{sec: Achievabilty 1}
In this section, we present an achievable scheme satisfying the privacy and correctness conditions for any $M \in \mathbb{N}$ and any $R_k\in \mathbb{R^{+}} \cup \{0\},\forall k \in [K]$, subject to \eqref{2-6-1} and \eqref{2-7}.
First, we illustrate the intuition behind the achievable scheme by the first part of a running example. Then, we present a general algorithmic scheme for nonnegative integer rates and after that we will elaborate on the achievable scheme and its feasibility in Subsections \ref{The Achievable Algorithm} and \ref{Feasibility_1}, respectively. As the general algorithmic achievable scheme is unravelling we apply each step on the running example. The proof of correctness and privacy can be found in  Subsections \ref{Correctness_1} and \ref{Privacy_1}, respectively.
In Subsection \ref{Memory Sharing}, a memory-sharing approach is applied to provide a coherent achievable scheme for nonnegative and real rates.

\begin{example}\emph{{Introducing a Simple DMUSS System}}:\label{Example1}
Consider a  multi-user secret sharing system with a mater nodes, $K=4$ users, and  $N=8$ storage nodes, each with storage size of
$M=1$ symbol of $GF(11)$, where user $k \in [K]$ has access to  $\A_k$, as
%Let $K=4,N=8,q=11,M=1$ and the access sets are chosen arbitrarily as follows,
\begin{gather}
    \A_1=\{1,6,7,8\},\:\A_2=\{1,3,4,7\},\:\A_3=\{1,2,3,8\},\:\A_4=\{2,4,5,6,7\}.
\end{gather}
Theorem \ref{Theorem 1} states that the capacity region of this system consists of all rate tuples  $(R_1, R_2, R_3, R_4)$ satisfying \eqref{3-1} and  \eqref{3-2}, i.e.,
\begin{align*}
&R_1\leq\min_{i\neq 1}|\A_1\backslash\A_i|=2,\:    R_2\leq\min_{i\neq 2}|\A_2\backslash\A_i|=2,R_3\leq\min_{i\neq 3}|\A_3\backslash\A_i|=2,\:R_4\leq\min_{i\neq 4}|\A_4\backslash\A_i|=3\\
&R_1+R_2\leq6,\:    R_1+R_3\leq6,\:     R_1+R_4\leq7,\:      R_2+R_3\leq6,\:       R_2+R_4\leq7,\:        R_3+R_4\leq8,\:\notag\\
& R_1+R_2+R_3\leq7,\:                  R_2+R_3+R_4\leq8,\:                           R_1+R_2+R_4\leq8,\:                            R_1+R_3+R_4\leq8,\notag\\
& R_1+R_2+R_3+R_4\leq8.
\end{align*}
Thus, $R_1=1,R_2=2,R_3=2,R_4=3$ $q$-array satisfy Theorem~\ref{Theorem 1} and are achievable.
\begin{figure}
\centering
\includegraphics[scale=0.8]{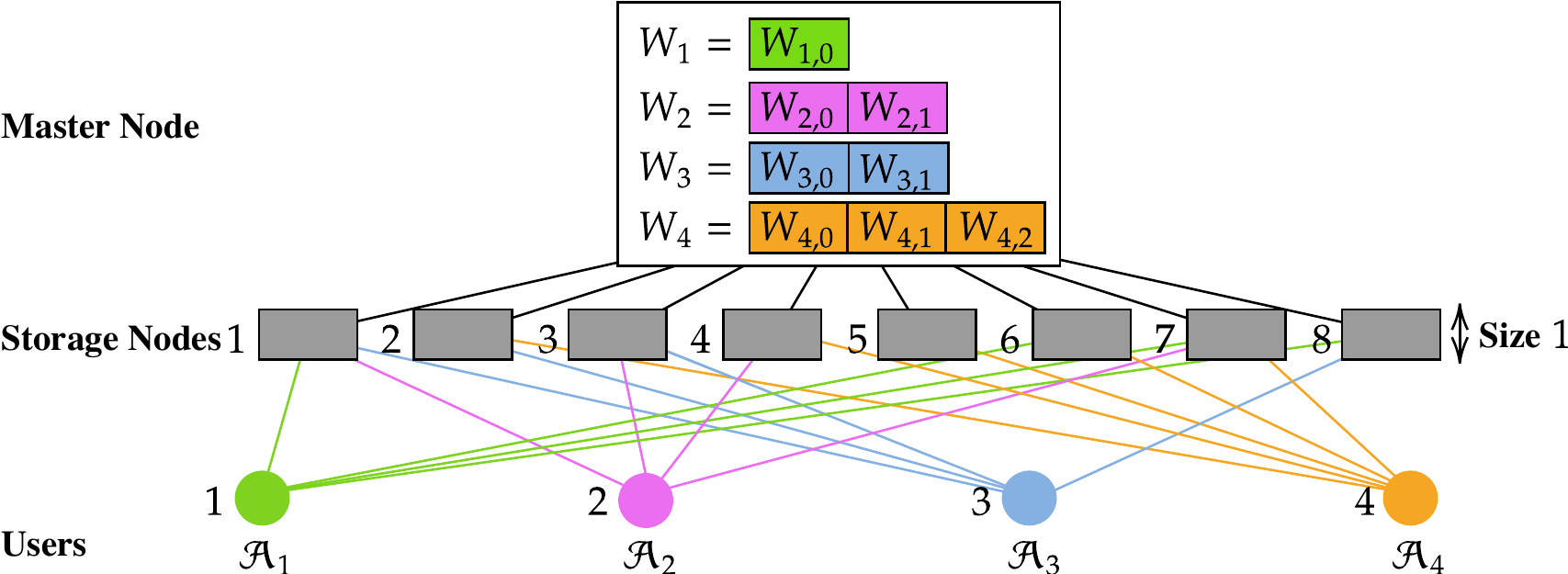}
\caption{Example 1: A DMUSS system consisting of a master node, $8$ storage nodes and $4$ users connected through some error-free links to a subset of the storage nodes. $\A_i,W_i,\: i\in [4],$ represents access set and secret message of user $i$, respectively.%According to Theorem \ref{Theorem 1}, $R_1=1,\:R_2,R_3=2,\: R_4=3$ are some achievable rates for the DMUSS system.
}\label{fig:2_1}
\end{figure}
\end{example}
\subsection{The Achievable Algorithm}
\label{The Achievable Algorithm}
In this section, we present the formal  description of the proposed algorithm. First, we introduce some  variables and  notations. Then,  we explain how the encoding scheme  determines the content of each storage node. After that,  we show how each user can retrieve its own secret message from the content of the storage nodes that it has access to.
In Subsections \ref{Feasibility_1}, \ref{Correctness_1}, and \ref{Privacy_1}, the feasibility, correctness, and privacy of the proposed scheme will be mathematically established, respectively.

Let us define $N\triangleq|\cup_{k\in[K]}\A_k|$ as the number of storage nodes.
 Suppose that the master node has $K$ secret messages with rates $(R_1,R_2,\hdots,R_K)\in(\mathbb{N} \cup \{0\})^K$,  satisfying \eqref{3-1} and \eqref{3-2} of Theorem \ref{Theorem 1}. Later we will show how to achieve noninteger data rates.
Thus by Definition~~\ref{2-1}, $W_k$ consists of $R_k$ symbols from $\F$ denoted by $W_{k,0},\hdots,W_{k,R_k-1}$ as shown in Fig. \ref{fig:1}.  We define $\mathbf{w}_k$ as
\begin{align}
    \mathbf{w}_k &\triangleq[W_{k,0},W_{k,1},\hdots,W_{k,{R_k-1}}]^\intercal, \: \forall k \in [K],\label{4-2}
\end{align}
as a  vector representation of message $W_k$. Recall that each user $k \in [K]$ has access to a subset of storage nodes $\mathcal{A}_k$.   We represent the elements in $\mathcal{A}_k$ by  $\{n_{k,1},\hdots,n_{k,|\A_k|}\}$.
We also  define
\begin{align}
        \mathbf{y} &\triangleq[Y_1,Y_2,\hdots,Y_N]^{\intercal},\label{4-1-2}
\end{align}
where $Y_n$ denotes the encoded symbol stored in storage node $n\in [N]$.  User $k$ only has access to $Y_{\A_k}= \{ Y_n, n \in \mathcal{A}_k \}$.

\subsubsection{Initialization}\label{init}
As the initialization phase,  the master nodes chooses some parameters, which are part of the protocol and shared with all users.  Apparently, these parameters are not functions of the secret messages.
The first set of parameters are denoted by $R'_k \in \mathbb{N} \cup \{0\}$, and are chosen such that\footnote{There exist such variables since the region is convex and $\sum^{K}_{k= 1}R'_k=N$ represents a surface that is always active since we know $|\A_{\S_1 \cup \S_2}| \leq |\A_{\S_1}|+|\A_{\S_2}|,\: \forall \S_1, \S_2 \subset [K]$.}
\begin{align}
  \sum_{k \in \S}R_k&\leq\sum_{k\in \S}R'_k\leq |\cup_{k \in \S} \A_k|,\: \forall \S \subset [K],\label{4-22-2222}\\
  \sum^{K}_{k=1}R_k&\leq\sum^{K}_{k= 1}R'_k=N.\label{4-22-222}
\end{align}
These variables will be used in  the description of the achievable algorithms in order to make it more readable.

The master node  also chooses some constants  $\alpha_{k,n} \in \mathbb{F}{\backslash\{0\}},\: \forall k\in[K], n\in [N]$ and $\gamma_{k,i} \in \mathbb{F}{\backslash\{0\}},\: \forall k\in[K],\: n_{k,i}\in \A_k$. For all $k\in[K]$, each $\gamma_{k,n}$ are distinct elements from $\mathbb{F}$ for any $n\in{\A_k}$. Note that these constants are not a function of the secret messages.
We define
\begin{align}
       \mathbf{ \boldsymbol{\alpha}}_k &\triangleq[ {\alpha}_{k,1},\alpha_{k,2},\hdots,\alpha_{k,{N}}]^{\intercal},\: \forall k\in [K],\label{4-1}\\
       \boldsymbol{\gamma}_k &\triangleq[\gamma_{k,1},\gamma_{k,2},\hdots,\gamma_{k,|\A_k|}]^{\intercal},\: \forall k \in [K],\label{4-1-1}
\end{align}
where $\alpha_{k,n} = 0, \forall k\in[K], n \notin \A_k$.
\subsubsection{Placement Phase}\label{Encoding}
The master nodes forms some vectors  $\mathbf{p}_k$, $k \in [K]$,  defined as
 \begin{align}
 \mathbf{p}_k &\triangleq[P_{k,R_k},P_{k,R_k+1},\hdots,P_{k,|\A_k|-1}]^\intercal \in \F^{(|\A_k|-R_k) \times 1}.
      \label{4-3-1-1}
\end{align}
Some of the entries of the above vectors are chosen randomly and the rest are calculated through solving some linear equations.
In particular, the master node  chooses  $P_{k,R_k},\hdots,P_{k,R'_k-1},\: \forall k \in [K]$,  independently and uniformly at random from $\F$, and also independently from the secret messages.
Then the master node forms the following set of equations:
\begin{align}
    g_{\mathbf{w}_k,\mathbf{p}_k}(\gamma_{k,i})&=-\alpha_{k,n_{k,i}}Y_{n_{k,i}},\quad\quad k \in [K],\: n_{k,i} \in \A_k,
      \label{4-3-1}
\end{align}
which can be expanded as
\begin{align}
W_{k,0} +\hdots+W_{k,R_k-1}\gamma_{k,i}^{R_k-1}     +P_{k,R_k}\gamma_{k,i}^{R_k}+&\hdots+P_{k,{|\A_k|-1}}\gamma_{k,i}^{{|\A_k|}-1} =-\alpha_{k,n_{k,i}}Y_{n_{k,i}},\label{4-3-1-1-1}
\end{align}
for $\forall k\in [K],\: n_{k,i}\in \A_k$.
In the set of linear equations \eqref{4-3-1-1-1},  $W_{k,0}, \ldots, W_{k,R_k-1}$, and $P_{k,R_k},\hdots,P_{k,R'_k-1}$,   $\forall k \in [K]$, are known, and $P_{k,R'_k},\hdots,P_{k,|\A_k|-1}$ and $Y_n$,  $\forall k \in [K]$ and $n \in [N]$ are unknown.

Afterwards, the master solves for $\mathbf{y}$ using the aforementioned set of equations \eqref{4-3-1-1-1} and stores resulted $Y_n,\: n\in [N]$ in the storage node $n$. One can easily check that  the number of equations is the same as the number of variables. Later in Subsection \ref{Feasibility_1} we will show that these set of equations are linearly independent, which guarantees  the existence of a unique solution for \eqref{4-3-1}.
\begin{example}\label{initilization, encoding}
\emph{Encoding:} Consider the example with $N=8$ storage nodes and $K=4$ users with access structure
\begin{align*}
    \A_1=&\{n_{1,1}=1,\:n_{1,2}=6,\:n_{1,3}=7,\:n_{1,4}=8\},\A_2=\{n_{2,1}=1,\:n_{2,2}=3,\:n_{2,3}=4,\:n_{2,4}=7\},\\
    \A_3=&\{n_{3,1}=1,\:n_{3,2}=2,\:n_{3,3}=3,\:n_{3,4}=8\},\A_4=\{n_{4,1}=2,\:n_{4,2}=4,\:n_{4,3}=5,\:n_{4,4}=6,\:n_{4,5}=7\}.
\end{align*}
It is easy to see that the rate tuples  $(R_1, R_2, R_3, R_4)=(1,2,2,3)$ satisfies  \eqref{3-1} and \eqref{3-2}. We have,
\begin{align*}
     \boldsymbol{\alpha}_1 &=[\alpha_{1,1},0,0,0,0,\alpha_{1,6},\alpha_{1,7},\alpha_{1,8}]^\intercal \:,\:     \boldsymbol{\alpha}_2 =[\alpha_{2,1},0,\alpha_{2,3},\alpha_{2,4},0,0,\alpha_{2,7},0]^\intercal,\\
     \boldsymbol{\alpha}_3 &=[\alpha_{3,1},\alpha_{3,2},\alpha_{3,3},0,0,0,0,\alpha_{3,8}]^\intercal  \:,\:
     \boldsymbol{\alpha}_4 =[0,\alpha_{4,2},0,\alpha_{4,4},\alpha_{4,5},\alpha_{4,6},\alpha_{4,7},0]^\intercal\\
    \boldsymbol{\gamma}_1&=[\gamma_{1,1},\gamma_{1,2},\gamma_{1,3},\gamma_{1,4}]^\intercal   \:,\:  \boldsymbol{\gamma}_2=[\gamma_{2,1},\gamma_{2,2},\gamma_{2,3},\gamma_{2,4}]^\intercal\\
      \boldsymbol{\gamma}_3&=[\gamma_{3,1},\gamma_{3,2},\gamma_{3,3},\gamma_{3,4}]^\intercal  \:,\: \boldsymbol{\gamma}_4=[\gamma_{4,1},\gamma_{4,2},\gamma_{4,3},\gamma_{4,4},\gamma_{4,5}]^\intercal
\end{align*}
Since $\sum^{4}_{i=1} R_i=8$, the master asserts $R'_i=R_i,\: \forall i\in [4]$.
Suppose that the master nodes has the secret messages $ \mathbf{w}_1=[1]^\intercal$, $\mathbf{w}_2=[2,6]^\intercal$, $\mathbf{w}_3=[4,0]^\intercal$, $\mathbf{w}_4=[3,5,7]^\intercal$.
For encoding, the master node forms the vectors $\mathbf{p}_1 =[P_{1,1}, P_{1,2},P_{1,3}]^\intercal$, $\mathbf{p}_2 =[P_{2,2}, P_{2,3}]^\intercal$,
$\mathbf{p}_3 =[P_{3,3}, P_{3,3}]^\intercal$, $\mathbf{p}_4 =[P_{4,1}, P_{4,4}]^\intercal$.  Since $R'_k = R_k,\: \forall k \in [K]$,  none of the entries of these vectors are chosen randomly. Then the maser node forms the following set of equations. For User 1,
\begin{align}
1 + P_{1,1}\gamma^1_{1,1}+P_{1,2}\gamma^2_{1,1}+P_{1,3}\gamma^3_{1,1}&= -\alpha_{1,1} Y_1 \:\:,\:\:%\label{ex-30}\\
1 + P_{1,1}\gamma^1_{1,2}+P_{1,2}\gamma^2_{1,2}+P_{1,3}\gamma^3_{1,2}= -\alpha_{1,6} Y_6,\label{ex-31}\\
1 + P_{1,1}\gamma^1_{1,3}+P_{1,2}\gamma^2_{1,3}+P_{1,3}\gamma^3_{1,3}&= -\alpha_{1,7} Y_7\:\:,\:\:%\label{ex-32}\\
1 + P_{1,1}\gamma^1_{1,4}+P_{1,2}\gamma^2_{1,4}+P_{1,3}\gamma^3_{1,4}= -\alpha_{1,8} Y_8.\label{ex-33}
\end{align}
Accordingly for User $2$, we have
\begin{align}
2 + 6\gamma^1_{2,1}+P_{2,2}\gamma^2_{2,1}+P_{2,3}\gamma^3_{3,1}&= -\alpha_{2,1} Y_1\:\:,\:\:%\label{ex-34}\\
2 + 6\gamma^1_{2,2}+P_{2,2}\gamma^2_{2,2}+P_{2,3}\gamma^3_{3,2}= -\alpha_{2,3} Y_3,\label{ex-35}\\
2 + 6\gamma^1_{2,3}+P_{2,2}\gamma^2_{2,3}+P_{2,3}\gamma^3_{3,3}&= -\alpha_{2,4} Y_4\:\:,\:\:%\label{ex-36}\\
2 + 6\gamma^1_{2,4}+P_{2,2}\gamma^2_{2,4}+P_{2,3}\gamma^3_{3,4}= -\alpha_{2,7} Y_7.\label{ex-37}
\end{align}
Similarly, for User $3$, we have
\begin{align}
4 +P_{3,2}\gamma^2_{3,1}+P_{3,3}\gamma^3_{3,1}&= -\alpha_{3,1} Y_1\:\:,\:\:%\label{ex-38}\\
4 +P_{3,2}\gamma^2_{3,2}+P_{3,3}\gamma^3_{3,2}= -\alpha_{3,2} Y_2,\label{ex-39}\\
4 +P_{3,2}\gamma^2_{3,3}+P_{3,3}\gamma^3_{3,3}&= -\alpha_{3,3} Y_3\:\:,\:\:%\label{ex-40}\\
4 +P_{3,2}\gamma^2_{3,4}+P_{3,3}\gamma^3_{3,4}= -\alpha_{3,8} Y_8
.\label{ex-41}
\end{align}
Finally for User $4$, we have
\begin{align}
3 + 5\gamma^1_{4,1}+7\gamma^2_{4,1}+P_{4,3}\gamma^3_{4,1}+P_{4,4}\gamma^4_{4,1}&= -\alpha_{4,2} Y_2\label{ex-42}\\
3 + 5\gamma^1_{4,2}+7\gamma^2_{4,2}+P_{4,3}\gamma^3_{4,2}+P_{4,4}\gamma^4_{4,2}&= -\alpha_{4,4} Y_4,\label{ex-43}\\
3 + 5\gamma^1_{4,3}+7\gamma^2_{4,3}+P_{4,3}\gamma^3_{4,3}+P_{4,4}\gamma^4_{4,3}&= -\alpha_{4,5} Y_5,\label{ex-44}\\
3 + 5\gamma^1_{4,4}+7\gamma^2_{4,4}+P_{4,3}\gamma^3_{4,4}+P_{4,4}\gamma^4_{4,4}&= -\alpha_{4,6} Y_6,\label{ex-45}\\
3 + 5\gamma^1_{4,5}+7\gamma^2_{4,5}+P_{4,3}\gamma^3_{4,5}+P_{4,4}\gamma^4_{4,5}&= -\alpha_{4,7} Y_7.\label{ex-46}
\end{align}
Note that \eqref{ex-31}-\eqref{ex-46} construct a system of linear equations, where $\mathbf{w}_k,\:k\in [4]$ are known variables, and  $\mathbf{p}_k, k \in [4]$ and $Y_n,\: n \in [8]$ are unknown variables. The total number of unknown variables is  $17$ which is exactly equal to the number of equations. The master node solves this set of linear equations and stores $Y_1$ to $Y_8$ in the storage nodes $1$ to $8$, respectively. Later in Subsection \ref{Feasibility_1}, we will prove that these equations are linearly independent, guaranteeing the existence of a unique solution.
\end{example}
\subsubsection{Retrieval Phase}
\label{Decoding}
Having access to the stored data $Y_{\A_k}$, User $k$ forms the following set of equations
\begin{align}
    g_{\hat{\mathbf{w}}_k,\hat{\mathbf{p}}_k}(\gamma_{k,i})&=-\alpha_{k,n_{k,i}}Y_{n_{k,i}},\:\forall n_{k,i} \in \A_k, \label{4-3-1-4}
\end{align}
where
\begin{align}
 \hat{\mathbf{w}}_k &\triangleq[\hat{W}_{k,0},\hat{W}_{k,1},\hdots,\hat{W}_{k,{R_k-1}}]^\intercal \in \F^{R_k \times 1}
\end{align}
and
\begin{align}
\hat{\mathbf{p}}_k \triangleq  [\hat{P}_{k,R_k},\hat{P}_{k,R_k+1},\hdots,\hat{P}_{k,|\A_k|-1}]^\intercal \in \F^{(|\A_k|-R_k) \times 1}
\end{align}
are unknown. After solving the set of equations \eqref{4-3-1-4}, User $k$ retrieves $\hat{\mathbf{w}}_k$ as a decoded version of $\mathbf{w}_k$.

\begin{example}\emph{Retrieval}: Each user forms a set of equations based on \eqref{4-3-1-4} by having its accessible storage nodes. For example, User $1$ has access to $Y_1,Y_6,Y_7,Y_8$ and  forms the following set of equations,
\begin{align}
\hat{W}_{1,0} + \hat{P}_{1,1}\gamma^1_{1,1}+\hat{P}_{1,2}\gamma^2_{1,1}+\hat{P}_{1,3}\gamma^3_{1,1}&= -\alpha_{1,1} Y_1,\label{ex-50}\\
\hat{W}_{1,0}  + \hat{P}_{1,1}\gamma^1_{1,2}+\hat{P}_{1,2}\gamma^2_{1,2}+\hat{P}_{1,3}\gamma^3_{1,2}&= -\alpha_{1,6} Y_6,\label{ex-51}\\
\hat{W}_{1,0} + \hat{P}_{1,1}\gamma^1_{1,3}+\hat{P}_{1,2}\gamma^2_{1,3}+\hat{P}_{1,3}\gamma^3_{1,3}&= -\alpha_{1,7} Y_7,\label{ex-52}\\
\hat{W}_{1,0}  + \hat{P}_{1,1}\gamma^1_{1,4}+\hat{P}_{1,2}\gamma^2_{1,4}+\hat{P}_{1,3}\gamma^3_{1,4}&= -\alpha_{1,8} Y_8.\label{ex-53}
\end{align}
User 1 derives $\hat{W}_{1,0}$, $\hat{P}_{1,1}$, $\hat{P}_{1,2}$, and $\hat{P}_{1,3}$ from the above system of equations.
Similar procedure is followed by other users.
\end{example}
We need to prove the feasibility and correctness of the placement and retrieval phases. In addition,  we need to  prove that in the proposed scheme  the privacy condition \eqref{2-7} is  satisfied.  In other words, we need to prove that there exist $\alpha_{k,n_i},\gamma_{k,i}\: k \in [K], n_i \in \A_k$, such that  following conditions hold:
\begin{enumerate}
    \item \emph{Feasibility:} The encoding set of equations \eqref{4-3-1} can be solved by the master node.
    \item \emph{Correctness:} The decoding set of equations \eqref{4-3-1-4} can be solved in such a way that $\hat{W}_k=W_k$.
    \item \emph{Privacy:}  The proposed solution satisfies \eqref{2-7}.
\end{enumerate}
The proof of the feasibility and correctness of the above encoding and decoding are presented in Subsections \ref{Feasibility_1} and \ref{Correctness_1}, respectively, and the proof of privacy is presented in Subsection \ref{Privacy_1}.

\subsection{Feasibility}\label{Feasibility_1}
In this subsection we prove that the  placement phase in Subsection~\ref{The Achievable Algorithm} is feasible.
%\\textbf{Formal Proof:}
%\subsubsection{Rewriting The Linear Equation}
First we write the set of encoding equations \eqref{4-3-1} in matrix form as follows,
\begin{equation}
\mathbf{A}\mathbf{b}=\mathbf{s},\label{4-7}
\end{equation}
where
\begin{align}
    \mathbf{b}&\triangleq[ \mathbf{p}_1(R'_1+1:|{\A}_K|)^\intercal,\hdots,\mathbf{p}_K(R'_K+1:|{\A}_K|)^\intercal,\mathbf{y}^\intercal]^\intercal,\label{4-7-1}
\end{align}
is the vector of unknown variables,
\begin{align}
\mathbf{s} &\triangleq [ \mathbf{s}_{1}^\intercal,\hdots,\mathbf{s}^\intercal_{K}]^\intercal,\label{4-8}\\
   \mathbf{s}_{k} &\triangleq - [ g_{\mathbf{w}_k}(\gamma_{k,1}),g_{\mathbf{w}_k}(\gamma_{k,2}),\hdots,g_{\mathbf{w}_k}(\gamma_{k,|\A_k|})]^\intercal,\: \forall k \in [K],\: n_{k,i} \in \A_k,    \label{4-8-1}
\end{align}
is the vector of known variables, and
  \begin{align}
\mathbf{A}_{}&\triangleq
\left[ \begin{array}{cccc|c}
  \mathbf{B}^\intercal_1& & & &\mathbf{\Upsilon}_1\\
   & \mathbf{B}^\intercal_2 & & &\mathbf{\Upsilon}_2\\
  & & \ddots& &\vdots\\
 & & & \mathbf{B}^\intercal_K &\mathbf{\Upsilon}_K\\
\end{array}\right]\: \in \F^{u \times u},\label{4-9}
\end{align}
is the matrix of coefficients, where $u=\sum^{K}_{k=1}|\mathcal{A}_k|$ and
 \begin{align}
\mathbf{B}_k &\triangleq
\begin{bmatrix}
\gamma_{k,1}^{R'_k } & \hdots & \gamma_{k,{|\A_k|}}^{R'_k }
\\\gamma_{k,1}^{R'_k +1} &\hdots &\gamma_{k,{|\A_k|}}^{R'_k +1}
\\\vdots&\ddots & \vdots
\\\gamma_{k,1}^{{|\A_k|}-1}&\hdots  & \gamma_{k,{|\A_k|}}^{{|\A_k|}-1}
\end{bmatrix},\: \forall k \in [K],
\label{4-10}\\
\mathbf{\Upsilon}_k   &\in \mathbb{F}^{|\A_k|\times N}.
\label{4-11}
\end{align}
All entries of  $\mathbf{\Upsilon}_k$ are $ 0 \in\mathbb{F}$, except   $\mathbf{\Upsilon}_k(i,n_{k,i}) =  \alpha_{k,n_{k,i}},  n_{k,i}\in \A_k$,$ i \in [|\A_k|]$.

\begin{example}\emph{Feasibility: rewriting encoding equations.}
Referring to Example Part  \ref{initilization, encoding},  we have
\begin{align}
    \mathbf{b}&=[ P_{1,1}, P_{1,2}, P_{1,3},P_{2,2},P_{2,3},P_{3,2},P_{3,3},P_{4,3},P_{4,4},\mathbf{y}^\intercal]^\intercal\;,\;%\label{ex-60}\\
    \mathbf{s} = [\mathbf{s}_1^\intercal,\mathbf{s}_2^\intercal,\mathbf{s}_3^\intercal,\mathbf{s}_4^\intercal] ^\intercal \label{ex-61}\\
    \mathbf{s}_1 &= -[1,1,1,1]^\intercal\;,\;%\label{ex-62}\\
    \mathbf{s}_2 = -[2 + 6\gamma^1_{2,1},2 + 6\gamma^1_{2,2},2 + 6\gamma^1_{2,3},2 + 6\gamma^1_{2,4}]^\intercal\;,\;%\label{ex-63}\\
     \mathbf{s}_3 = -[4,4,4,4]^\intercal\label{ex-64}\\
      \mathbf{s}_4 &=- [3 + 5\gamma^1_{4,1}+7\gamma^2_{4,1},
      3 + 5\gamma^1_{4,2}+7\gamma^2_{4,2},
      3 + 5\gamma^1_{4,3}+7\gamma^2_{4,3},
      3 + 5\gamma^1_{4,4}+7\gamma^2_{4,4}, 3 + 5\gamma^1_{4,5}+7\gamma^2_{4,5}]^\intercal.\label{ex-65}
      \end{align}
In addition, $\mathbf{A}$ is as \eqref{4-9} with $K=4$, and
$u=\sum^{4}_{i=1} |\A_i| = 17$. Also we have,
\begin{align}
\mathbf{B}^{}_1 &\triangleq
\begin{bmatrix}
\gamma_{1,1}^{1 } & \gamma_{1,2}^{1 } & \gamma_{1,3}^{1 }& \gamma_{1,{4}}^{1}
\\
\gamma_{1,1}^{2 } & \gamma_{1,2}^{2 } & \gamma_{1,3}^{2 }& \gamma_{1,{4}}^{2}
\\\gamma_{1,1}^{3 } & \gamma_{1,2}^{3} & \gamma_{1,3}^{3 }& \gamma_{1,{4}}^{3}
\end{bmatrix}
\;,\;%\label{ex-66}\\
\mathbf{B}_2 \triangleq
\begin{bmatrix}
\gamma_{2,1}^{2 } & \gamma_{2,2}^{2 } & \gamma_{2,3}^{2 }& \gamma_{2,{4}}^{2}
\\\gamma_{2,1}^{3 } & \gamma_{2,2}^{3} & \gamma_{2,3}^{3 }& \gamma_{2,{4}}^{3}
\end{bmatrix},
\label{ex-67}\\
\mathbf{B}_3 &\triangleq
\begin{bmatrix}
\gamma_{3,1}^{2 } & \gamma_{3,2}^{2 } & \gamma_{3,3}^{2 }& \gamma_{3,{4}}^{2}
\\\gamma_{3,1}^{3 } & \gamma_{3,2}^{3} & \gamma_{3,3}^{3 }& \gamma_{3,{4}}^{3}
\end{bmatrix}
\;,\;%\label{ex-68}\\
\mathbf{B}_4 \triangleq
\begin{bmatrix}
\gamma_{4,1}^{3 } & \gamma_{4,2}^{3 } & \gamma_{4,3}^{3 }& \gamma_{4,{4}}^{3} & \gamma_{4,{5}}^{3}
\\\gamma_{4,1}^{4 } & \gamma_{4,2}^{4} & \gamma_{4,3}^{4 }& \gamma_{4,{4}}^{4} &\gamma_{4,{5}}^{4}
\end{bmatrix},\label{ex-69}
\end{align}
\begin{align}
\boldsymbol{\Upsilon}_1 &\triangleq
\begin{bmatrix}
\alpha_{1,1}& 0 & 0 &0& 0 &  0 &0& 0\\
0& 0 & 0 &0& 0 & \alpha_{1,6} &0& 0\\
0& 0 & 0 &0& 0 & 0 &\alpha_{1,7}& 0\\
0& 0 & 0 &0& 0 & 0 &0& \alpha_{1,8}\\
\end{bmatrix}\;,\;%\label{ex-70}\\
\boldsymbol{\Upsilon}_2 \triangleq
\begin{bmatrix}
\alpha_{2,1}& 0 & 0 &0& 0 &  0 &0& 0\\
0& 0 & \alpha_{2,3}&0& 0 & 0 &0& 0\\
0& 0 & 0 &\alpha_{2,4}& 0 & 0 &0& 0\\
0& 0 & 0 &0& 0 & 0 &\alpha_{2,7}& 0\\
\end{bmatrix},\label{ex-71}\\
\boldsymbol{\Upsilon}_3 &\triangleq
\begin{bmatrix}
\alpha_{3,1}& 0 & 0 &0& 0 &  0 &0& 0\\
0& \alpha_{3,2} & 0&0& 0 & 0 &0& 0\\
0& 0 &\alpha_{3,3} &0& 0 & 0 &0& 0\\
0& 0 & 0 &0& 0 & 0 &0& \alpha_{3,8}\\
\end{bmatrix}\;,\;%\label{ex-72}\\
\boldsymbol{\Upsilon}_4 \triangleq
\begin{bmatrix}
0& \alpha_{4,2} & 0 &0& 0 &  0 &0& 0\\
0& 0 & 0& \alpha_{4,4}& 0 & 0 &0& 0\\
0& 0 &0 &0& \alpha_{4,5} & 0 &0& 0\\
0& 0 & 0 &0& 0 & \alpha_{4,6} &0& 0\\
0& 0 & 0 &0& 0 & 0 &\alpha_{4,7}& 0
\end{bmatrix}.\label{ex-72}
\end{align}
\end{example}

 A  solution for \eqref{4-7} exists,  if $\mathbf{A}$ is a nonsingular matrix.

\begin{lemma}\label{lemma1}
Assume  $|\F| > \max_{i \in [K]} |\A_i|$. There exist $\alpha_{k,n_{k,i}},\gamma_{k,i} \in \mathbb{F}\backslash \{0\}, k \in[K],\:n_{k,i}\in \A_k$,  such that matrix $\mathbf{A}$, defined in \eqref{4-9},   is nonsingular.
\end{lemma}
To prove the lemma,  it is enough  to prove that there exist $\alpha_{k,n_{k,i}},\gamma_{k,i} \in \mathbb{F}\backslash \{0\}, k \in[K],\:n_{k,i}\in \A_k$ such that  if
\begin{equation}
\mathbf{A}^\intercal \boldsymbol{\lambda}=\mathbf{0}
,\label{eq:5-1}
\end{equation}
then
\begin{equation}
\boldsymbol{\lambda}=\mathbf{0}.\label{eq:5-1-3}
\end{equation}
Let us partition $\boldsymbol{\lambda}$ as
\begin{equation}
\boldsymbol{\lambda} \triangleq [\boldsymbol{\lambda}_1^\intercal, \boldsymbol{\lambda}_2^\intercal, \hdots,\boldsymbol{\lambda }_K^\intercal]^\intercal,
\label{eq:5-1-1}
\end{equation}
where
\begin{equation}
\boldsymbol{\lambda}_{k} \in \F^{|\A_k|\times1},\:\: \forall k \in [K].\label{eq:5-3}
\end{equation}
 From \eqref{4-9} and \eqref{eq:5-1-1}, we conclude that  \eqref{eq:5-1} is equivalent to
\begin{equation}
\mathbf{B}_k\boldsymbol{\lambda}_k=\mathbf{0},\: \forall k \in [K], \label{eq:5-2}
\end{equation}
\begin{equation}
\mathbf{\Upsilon}_1^\intercal\boldsymbol{\lambda}_1+\mathbf{\Upsilon}_2^\intercal\boldsymbol{\lambda}_2+\hdots+\mathbf{\Upsilon}_K^\intercal\boldsymbol{\lambda}_K=0.
 \label{eq:5_9}
\end{equation}
{Note that system of equations mentioned in \eqref{eq:5_9} is related to the access sets. }

{From now on, our goal is to prove that $\boldsymbol{\lambda}_k,k\in [K]$, are zero vectors by utilizing \eqref{eq:5-2} and \eqref{eq:5_9}. }
We choose
\begin{align}
   \gamma_{k,i}=\gamma^{\pi_{k}(i)},\forall k\in[K],n_{k,i}\in \A_k, \label{eq:5-5-1}
\end{align}
where $\gamma$ is a primitive element in $\mathbb{F}$ and
\begin{equation}
    \pi_{k}{}: [|\A_k|] \xrightarrow{} [|\A_k|],
    \label{eq:5-5}
\end{equation}
 is a \emph{permutation function}. In what follows, we will prove that there are permutations $\pi_k,k \in [K]$, and $\alpha_{k,n_{k,i}} \in \mathbb{F}\backslash \{0\}, k \in[K],\:n_{k,i}\in \A_k$ such that $\mathbf{A}$ is nonsingular.  $\mathbf{B}_k^{(\pi_k)}$ denotes
 $\mathbf{B}_k$, with the particular choices of \eqref{eq:5-5-1} for  $\gamma_{k,i}$.

For some integers $m$ and $n$, we define $\mathbf{B}(m,n,\gamma)$ as
\begin{equation}
    \mathbf{B}(m,n,\gamma) \triangleq \begin{bmatrix}
(\gamma^{m })^{1 }&(\gamma^{m })^{{2 }}&\hdots&(\gamma^{{m }})^{{n}}\\
(\gamma^{m +1})^{ 1}&(\gamma^{m +1})^{{2 }}&\hdots&(\gamma^{{m +1}})^{{n}}\\
(\gamma^{m +2})^{1 }&(\gamma^{m +2})^{{2 }}&\hdots&(\gamma^{{m +2}})^{n}\\
\vdots&\vdots&\ddots&\vdots\\
(\gamma^{n-1})^{ 1}&(\gamma^{n-1})^{{2 }}&\hdots&(\gamma^{{n-1}})^{{n}}\\
\end{bmatrix},
\label{eq:5-7-1}
\end{equation}
then it is easy to see that
\begin{equation}
    \mathbf{B}^{(\pi_k)}_k = \mathbf{B}(R'_k,|\A_k|,\gamma) \mathbf{\Pi}_k,  \label{eq:5-555}
\end{equation}
where  matrix $\mathbf{\Pi}_k$ of size $|\A_k|\times |\A_k|$  is a permutation matrix associated with $\pi_k$ in which
{\begin{equation}
   \mathbf{\Pi}_k(i,\pi(i)) = 1,\: i \in [|\A_k|],
    \label{eq:5-7-2}
\end{equation}}
and other entries are zero.

Note that since $|\F| > \max_{k \in [K]} |\A_k|$ and $\gamma$ is a primitive element of $\F$, the entries of the first column of $\mathbf{B}(R'_k,|\A_k|,\gamma)$ are different, therefore $\mathbf{B}(R'_k,|\A_k|,\gamma)$  is a  Vandermonde matrix and is full rank. Thus, $\mathbf{B}^{(\pi_k)}_k$ is also full rank, for any $k \in [K]$.
From \emph{Rank–Nullity Theorem} \cite{r10},  the  dimension of the nullspace of $\mathbf{B}(R'_k,|\A_k|,\gamma)$ is equal to the number of columns minus the number of rows \emph{i.e.} $|\A_k|- (|\A_k|-R'_k)=R'_k$.
We denote the basis vectors of the null-space of $\mathbf{B}(R'_k,|\A_k|,\gamma)$  by  ${\mathbf{v}}^{(k)}_{1}, \ldots,  {\mathbf{v}}^{(k)}_{R'_k}$.
Let us define
\begin{equation}
    {\mathbf{v}}^{(k,\pi_{k})}_{r}
  \triangleq
 \Pi^{\intercal}_k  {\mathbf{v}}^{(k)}_{r}
 ,\: \forall r \in [R'_k].\label{eq:5_10_12}
\end{equation}
Since  $\mathbf{\Pi}_k$ is a unitary matrix and due to \eqref{eq:5-555},  $\mathbf{v}^{(k,\pi_{k})}_{1},\mathbf{v}^{(k,\pi_{k})}_2,\hdots,\mathbf{v}^{(k,\pi_{k})}_{R'_k}$ are linearly independent and also there are in the null-space of ${\mathbf{B}}^{(\pi_k)}_k$. Thus $\mathbf{v}^{(k,\pi_{k})}_{1},\mathbf{v}^{(k,\pi_{k})}_2,\hdots,\mathbf{v}^{(k,\pi_{k})}_{R'_k}$  forms the basis vectors of the null-space of ${\mathbf{B}}^{(\pi_k)}_k$.
Let us define
\begin{align}
    {\mathbf{V}}^{(k)}    &\triangleq [ {\mathbf{v}}^{(k)}_{1},  {\mathbf{v}}^{(k)}_{2}, \ldots,  {\mathbf{v}}^{(k)}_{R'_{k}}],\label{eq:V_k}\\
 {\mathbf{V}}^{(k,\pi_{k})}    &\triangleq [ {\mathbf{v}}^{(k,\pi_{k})}_{1},  {\mathbf{v}}^{(k,\pi_{k})}_{2}, \ldots,  {\mathbf{v}}^{(k,\pi_{k})}_{R'_{k}}].\label{eq:V_k}
\end{align}
Regarding \eqref{eq:5-2},   $ \boldsymbol{{\lambda}}_k $ is in the null-space of ${\mathbf{B}}^{(\pi_k)}_k$ and can be written as
\begin{equation}
   \boldsymbol{{\lambda}}_k = {\mathbf{V}}^{(k,\pi_{k})} \boldsymbol{\eta}_k, \label{eq:5_10_10}
\end{equation}
for some vector of scalars coefficients $\boldsymbol{\eta}_k \in \mathbb{F}^{R'_k}$.

\begin{example}\emph{Basis Vectors of Null-Space of $B_k$:} From \eqref{eq:5-5-1}, we know that the master node has assigned  $\gamma_{k,i}=8^{\pi_k(i)},\: \forall k \in [4],\: n_{k,i} \in \A_k$. Now, it has to determine the null-space of $ \mathbf{B}^{(\pi_k)}_k,\: k\in [4]$. Since $\pi_k, \:k \in [4]$ is yet to be determined, referring to \eqref{eq:5-555} and \eqref{eq:5_10_12} it is only needed to determine the null-space of $\mathbf{B}(R'_k,|\A_k|,\gamma), \forall k \in [4] $ and then construct the basis vectors of null-space of $\mathbf{B}^{(\pi_k)}_k,\: k\in [4]$ from the null-space of $\mathbf{B}(R'_k,|\A_k|,\gamma)$. Thus, for User $1$, we have
\begin{align}
    \mathbf{B}(R'_1,|\A_1|,\gamma) = \mathbf{B}(1,4,8)= \begin{bmatrix}
8^{1 } & 8^{2 } & 8^{3 }&8^{4}
\\
(8^{2 })^{1} & (8^{2 })^{2} & (8^{2 })^{3}& (8^{2 })^{4}
\\
(8^{3 })^{1}& (8^{3 })^{2} & (8^{3 })^{3}& (8^{3 })^{4}
\end{bmatrix},\label{ex-80}
\end{align}
and in this regards its null space is one dimensional and can be spanned by the following vector,
\begin{align}
    \mathbf{V}^{(1)} = [1,8,4,7]^\intercal\label{ex-81}.
\end{align}
Similarly for Users $2$ and $3$ we have,
\begin{align}
    \mathbf{B}(R'_2,|\A_2|,\gamma) = \mathbf{B}(R'_3,|\A_3|,\gamma) =
    \mathbf{B}(2,4,8) =
    \begin{bmatrix}
(8^{2 })^{1} & (8^{2 })^{2} & (8^{2 })^{3}& (8^{2 })^{4}
\\
(8^{3 })^{1}& (8^{3 })^{2} & (8^{3 })^{3}& (8^{3 })^{4}
\end{bmatrix}\label{ex-82},
\end{align}
and basis vectors of the null-space for $k=2,3$ are
\begin{align}
    \mathbf{V}^{(2)} = \mathbf{V}^{(3)} &= \begin{bmatrix}
1 & 8&4&7
\\
1&7&0&8
\end{bmatrix}^\intercal\label{ex-83}.
\end{align}
Finally, for User $4$ we have,
\begin{align}
    \mathbf{B}(R'_4,|\A_4|,\gamma) = \mathbf{B}(3,5,8) = \begin{bmatrix}
(8^{3 })^{1}& (8^{3 })^{2} & (8^{3 })^{3}& (8^{3 })^{4} &(8^3)^{5}\\
(8^{4 })^{1}& (8^{4 })^{2} & (8^{4 })^{3}& (8^{4 })^{4} &(8^4)^{5}\\
\end{bmatrix}\label{ex-85},
\end{align}
and basis vectors of the null-space for $k=4$ are
\begin{align}
\mathbf{V}^{(4)} &= \begin{bmatrix}
1& 5& 2& 6& 1\\
1& 6& 3& 4& 4\\
1& 1& 1& 10& 7
\end{bmatrix}^\intercal.\label{ex-86}
\end{align}
\end{example}

\subsubsection{Expansion in Terms of Basis Vectors of the Null-spaces}
 Substituting  \eqref{eq:5_10_10} in \eqref{eq:5_9}, we have
\begin{equation}
\sum^{K}_{k=1}\mathbf{\Upsilon}_k^\intercal  {\mathbf{V}}^{(k,\pi_{k})} \boldsymbol{\eta}_k=\boldsymbol{0}
 .\label{eq:5_11_5}
\end{equation}

We define
\begin{align}
    \boldsymbol{\eta} & \triangleq[\boldsymbol{\eta}_{1}^\intercal,\boldsymbol{\eta}_{2}^\intercal,\hdots,\boldsymbol{\eta}_{K}^\intercal]^\intercal,\label{eq:5_11_22}\\
    \boldsymbol{\Psi}_k^{(\pi_k)} & \triangleq \mathbf{\Upsilon}_k^\intercal  {\mathbf{V}}^{(k,\pi_{k})} ,\label{eq:5_12_12_1}\\
    \mathbf{V}^{(\pi)}& \triangleq  \big[\boldsymbol{\Psi}_1^{(\pi_1)}, \ldots, \boldsymbol{\Psi}_K^{(\pi_K)} ].   \label{matrixV}
\end{align}
Note that $\mathbf{V}^{(\pi)} \in \F ^{N \times \sum^{K}_{k=1}R'_k}$ is a square matrix since  \eqref{4-22-222} holds.
Then, we can rewrite \eqref{eq:5_11_5} as
\begin{equation}
\mathbf{V}^{(\pi)}\boldsymbol{\eta}=\mathbf{0}.
 \label{eq:5_12}
\end{equation}

Now our goal is to show  that there exist $\alpha_{k,n_{k,i}} \in \mathbb{F}\backslash \{0\}, k \in[K],\:n_{k,i}\in \A_k$ parameters and $\pi_k, k \in [K]$, permutations such that \eqref{eq:5_12} guarantees that $\boldsymbol{\eta}=0$, or equivalently  $\mathbf{V}^{(\pi)}$ is nonsingular.

To explain $\mathbf{V}^{(\pi)}$ in details, notice
\begin{equation}
\boldsymbol{\Psi}_k^{(\pi_k)}(\{n_{k,i}\},[1:R'_k])=
\begin{bmatrix}
\alpha_{k,n_{k,i}}v_{i,1} \:\: \alpha_{k,n_{k,i}}v_{i,2} \:\: \hdots \:\: \alpha_{k,n_{k,i}}v_{i,R'_k}
\end{bmatrix},
    \label{eq:5_12_12_2}
\end{equation}
and all other entries of $\boldsymbol{\Psi}_k^{(\pi_k)}$ would be zero.

\begin{example}\label{Example7}\emph{Feasibility: Constructing $\mathbf{V}^{(\pi)}$.}Referring to \eqref{eq:5_12}, we have

{\small\begin{equation*}
\begin{matrix}
\mathbf{V^{(\pi)}}
 =
 \begin{bmatrix}
 \bovermat{$\boldsymbol{\Psi}^{(\pi_1)}_1$}{ \alpha_{1,1}v_{1,1}^{(1,\pi_{1})}} & \bovermat{$\boldsymbol{\Psi}^{(\pi_2)}_2$}{ \alpha_{2,1}v_{1,1}^{(2,\pi_{2})} &\alpha_{2,1}v_{1,2}^{(2,\pi_{2})}}&\bovermat{$\boldsymbol{\Psi}^{(\pi_3)}_3$}{ \alpha_{3,1}v_{1,1}^{(3,\pi_{3})}  &
\alpha_{3,1}v_{1,2}^{(3,\pi_{3})}} & \bovermat{$\boldsymbol{\Psi}^{(\pi_4)}_4$}{\:\:\:\:\:\:0\:\:\:\:\:\:\:\:\:\:\:\:&\:\:\:\:\:\:0\:\:\:\:\:\:\:\:\: &\:\:\:\:\:\:\:\:\:\:\:\:0\:\:\:\:\:\:\:\:\:
}\\[0.5em]

%1
 0 &0 &0& \alpha_{3,2}v_{2,1}^{(3,\pi_{3})}  & \alpha_{3,2}v_{2,2}^{(3,\pi_{3})}&  \alpha_{4,2}v_{1,1}^{(4,\pi_{4})} &\alpha_{4,2}v_{1,2}^{(4,\pi_{4})} &\alpha_{4,2}v_{1,3}^{(4,\pi_{4})} \\[0.5em]
%
%2
 0 & \alpha_{2,3}v_{2,1}^{(2,\pi_{2})} & \alpha_{2,3}v_{2,2}^{(2,\pi_{2})} & \alpha_{3,3}v_{3,1}^{(3,\pi_{3})} & \alpha_{3,3}v_{3,2}^{(3,\pi_{3})}&0 & 0 &0^{
} \\[0.5em]
%3
 0 & \alpha_{2,4}v_{3,1}^{(2,\pi_{2})} & \alpha_{2,4}v_{3,2}^{(2,\pi_{2})} & 0 & 0  & \alpha_{4,4}v_{2,1}^{(4,\pi_{4})} &\alpha_{4,4}v_{2,2}^{(4,\pi_{4})} &\alpha_{4,4}v_{2,3}^{(4,\pi_{4})} \\[0.5em]
 0& 0&0 &0 &0  & \alpha_{4,5}v_{3,1}^{(4,\pi_{4})} &\alpha_{4,5}v_{3,2}^{(4,\pi_{4})} &\alpha_{4,5}v_{3,3}^{(4,\pi_{4})} \\[0.5em]
 \alpha_{1,6}v_{2,1}^{(1,\pi_{1})} & 0 &0& 0 &0  &\alpha_{4,6}v_{4,1}^{(4,\pi_{4})} &\alpha_{4,6}v_{4,2}^{(4,\pi_{4})} &\alpha_{4,6}v_{4,3}^{(4,\pi_{4})} \\[0.5em]
\alpha_{1,7}v_{3,1}^{(1,\pi_{1})}& \alpha_{2,7}v_{4,1}^{(2,\pi_{2})} & \alpha_{2,7}v_{4,2}^{(2,\pi_{2})}& 0 &0  &\alpha_{4,7}v_{5,1}^{(4,\pi_{4})} &\alpha_{4,7}v_{5,2}^{(4,\pi_{4})} &\alpha_{4,7}v_{5,3}^{(4,\pi_{4})} \\[0.5em]
\alpha_{1,8}v_{4,1}^{(1,\pi_{1})}& 0 &0&\alpha_{3,8}v_{3,1}^{(3,\pi_{3})} & \alpha_{3,8}v_{3,2}^{(3,\pi_{3})} &0&0& 0 \\[0.5em]
  \end{bmatrix}%
 \end{matrix}%\label{ex-90}
\end{equation*}}
\end{example}

\subsubsection{Nonsingularity of $\mathbf{V}^{(\pi)}$}
Now we are to prove $\mathbf{V}^{(\pi)}$ is  nonsingular, \emph{i.e.},
\begin{equation}
\textrm{rank}(\mathbf{V}^{(\pi)})=N.
\label{eq:5_16}
\end{equation}
For this purpose,  consider the sub-vector  $\hat{\boldsymbol{\alpha}_k}$,  consisting of only nonzero elements of $\boldsymbol{\alpha}_k$, defined as:
\begin{equation}
    \hat{\boldsymbol{\alpha}_k} = \boldsymbol{\alpha}_k(\A_k).
\end{equation}
 In addition we define $\hat{\mathbf{\Psi}}^{(\pi_k)}_{k}$  of dimension $|\A_k| \times R'_k$ as
\begin{equation}
   \hat{\mathbf{\Psi}}^{(\pi_k)}_{k} = {\diag({\hat{\boldsymbol{\alpha}}}_k}) \begin{bmatrix}
    \mathbf{v}^{(k,\pi_{k})}_{1} \: \mathbf{v}^{(k,\pi_{k})}_{2} \: \hdots \: \mathbf{v}^{(k,\pi_{k})}_{R'_k}
    \end{bmatrix}.
\end{equation}
Note that $\hat{\mathbf{\Psi}}^{(\pi_k)}_{k}$ is a sub-matrix of $\mathbf{\Psi}^{(\pi_k)}_{k}$. In addition, assuming none of the entries of   $\hat{\boldsymbol{\alpha}_k}$ is zero, and since $\begin{bmatrix}
    \mathbf{v}^{(k,\pi_{k})}_{1} \: \mathbf{v}^{(k,\pi_{k})}_{2} \: \hdots \: \mathbf{v}^{(k,\pi_{k})}_{R'_k}
    \end{bmatrix}$ is full rank,  $\hat{\mathbf{\Psi}}^{(\pi_k)}_{k}$ is a full rank matrix with rank $R'_k$.  Therefore,  there exists at least a nonsingular square sub-matrix of dimension $R'_k \times R'_k$ of matrix  $\hat{\mathbf{\Psi}}^{(\pi_k)}_{k}$,  which is  also sub-matrix of $\mathbf{\Psi}^{(\pi_k)}_{k}$.  Let  $\mathcal{Z}^{(\pi_k)}_k \subset \A_k$, $|\mathcal{Z}^{(\pi_k)}_k|=R'_k$  such that  $\mathbf{\Psi}^{(\pi_k)}_{k} (\mathcal{Z}^{(\pi_k)}_k, [R'_k])$ is a nonsingular square sub-matrix of dimension $R'_k \times R'_k$ of matrix  $\mathbf{\Psi}^{(\pi_k)}_{k}$.

Now there is an important observation. By changing the permutation function $\pi_k$, the master node can change  $\mathcal{Z}^{(\pi_k)}_k$ to be any arbitrary subset  of size $R'_k$ of $\A_k$. This is because the  permutation function $\pi_k$ allows the master node to relabel the rows of  $\mathbf{\Psi}^{(\pi_k)}_{k}$  anyway that it wishes.

We claim that there exits some $K$ subsets $\mathcal{Z}^*_k \subset [N]$,   $k \in [K]$,  satisfying the following conditions:
\begin{align}
|\Z^*_k| &= R'_k, \label{coherent-1} \\
\Z^*_k & \subset \A_k, \label{coherent-2}\\
\Z^*_k \cap \Z^*_{k'} & = \emptyset, \forall k \neq k'. \label{coherent-3}
\end{align}
The proof  of existence for such subsets $\mathcal{Z}^*_k \subset [N]$, $k \in [K]$,  follows from  Hall's marriage Theorem~\cite{r13} and can be found in Appendix \ref{A_2}.

To continue, the master node chooses the permutation function $\pi^*_k$ such that
\begin{equation}
    \mathcal{Z}^{(\pi^*_k)}_k=\Z^*_k,\:\forall k \in [K].\label{coherent-4}
\end{equation}
We also denote the corresponding matrix $\mathbf{V}^{(\pi)}$, defined in~\eqref{matrixV}, with $\mathbf{V}^{(\pi^*)}$ after this choice.
The procedure has been illustrated in the following example.
\begin{example}\emph{Permutation Assignment Effect}:
Let us first, determine the nonsingular sub-matrices of the basis vectors, mentioned  in \eqref{ex-81}, \eqref{ex-83} and \eqref{ex-86}. The red elements show nonsingular sub-matrices in the following statement,
\begin{equation}
     \mathbf{V}^{(1)} = \begin{bmatrix}
\color{red}{1}
\\
8
\\
4
\\
7
\end{bmatrix},\:
 \mathbf{V}^{(2)} = \mathbf{V}^{(3)} = \begin{bmatrix}
\color{red}{1} & \color{red}{1}
\\
\color{red}{8} &\color{red}{7}
\\
4 &0
\\
7 &8
\end{bmatrix},
\:
 \mathbf{V}^{(4)} = \begin{bmatrix}
\color{red}1 &\color{red} 1 &\color{red} 1
\\
\color{red}5 & \color{red}6 & \color{red}1
\\
\color{red}2 & \color{red}3 &\color{red} 1
\\
6 & 4 & 10
\\
1 & 4& 7
\end{bmatrix}.
\end{equation}
Also the master node determines $\Z^{*}_k,\: k \in [K],$ as follows,
\begin{equation}
    \Z^{*}_{1}= \{8\},\:  \Z^{*}_{2}= \{3,4\},\:  \Z^{*}_{3}= \{1,2\},\:  \Z^{*}_{4}= \{5,6,7\},
\end{equation}
In addition, the master node assigns permutation functions as follows,
\begin{align}
  \pi_1^{*} =
  \begin{pmatrix}
1 & 2 & 3 & 4\\
4 & 2 & 3 & 1
\end{pmatrix},
 \pi_2^{*} =
  \begin{pmatrix}
1 & 2 & 3 & 4\\
3 & 2 & 1 & 4
\end{pmatrix},
 \pi_3^{*} =
  \begin{pmatrix}
1 & 2 & 3 & 4\\
1 & 2 & 3 & 4
\end{pmatrix},
 \pi_4^{*} =
  \begin{pmatrix}
1 & 2 & 3 & 4 & 5\\
4 & 5 & 1 & 2 & 3
\end{pmatrix}.
\end{align}
Now the permuted basis vectors, defined in \eqref{eq:5_10_12}, are
\begin{align}
     \mathbf{V}^{(1,\pi_1)} = \begin{bmatrix}
7
\\
8
\\
4
\\
\color{red}{1}
\end{bmatrix},
 \mathbf{V}^{(2,\pi_2)} = \begin{bmatrix}
  4 &0 \\
  \color{red}{8} &\color{red}{7}
\\
\color{red}{1} & \color{red}{1}
\\
7 &8
\end{bmatrix},
 \mathbf{V}^{(3,\pi_3)} =
\begin{bmatrix}
\color{red}{1} & \color{red}{1}
\\
  \color{red}{8} &\color{red}{7}
\\
 4 &0 \\
7 &8
\end{bmatrix},
 \mathbf{V}^{(4,\pi_4)}&= \begin{bmatrix}
  6 & 4 & 10
\\
1 & 4& 7\\
\color{red}1 &\color{red} 1 &\color{red} 1
\\
\color{red}5 & \color{red}6 & \color{red}1
\\
\color{red}2 & \color{red}3 &\color{red} 1
\end{bmatrix}.
\end{align}
Therefore, $\mathbf{V}^{(\pi^{*})}$ becomes,

\begin{equation}
\begin{matrix}
\mathbf{V}^{(\pi^*)}
 =
 \begin{bmatrix}
 \bovermat{$\boldsymbol{\Psi}^{(\pi_1^{*})}_1$}{ \alpha_{1,1}7} & \bovermat{$\boldsymbol{\Psi}^{(\pi_2^{*})}_2$}{ \alpha_{2,1}4 &\alpha_{2,1} 0}&\bovermat{$\boldsymbol{\Psi}^{(\pi_3^{*})}_3$}{ \color{red} \alpha_{3,1}1  &
\color{red} \alpha_{3,1}1} & \bovermat{$\boldsymbol{\Psi}^{(\pi_4^{*})}_4$}{\:\:\:\:\:\:0\:\:\:\:\:\:\:\:\:\:\:\:&\:\:\:\:\:\:0\:\:\:\:\:\:\:\:\: &\:\:\:\:\:\:\:\:\:\:\:\:0\:\:\:\:\:\:\:\:\:
}\\[0.5em]

%1
 0 &0 &0& \color{red}\alpha_{3,2}8 & \color{red}\alpha_{3,2}7&  \alpha_{4,2}6 &\alpha_{4,2}4 &\alpha_{4,2}10 \\[0.5em]
%
%2
 0 & \color{red}\alpha_{2,3}8 & \color{red}\alpha_{2,3}7 & \alpha_{3,3}4& \alpha_{3,3}0&0 & 0 &0^{
} \\[0.5em]
%3
 0 & \color{red}\alpha_{2,4}1 & \color{red}\alpha_{2,4}1 & 0 & 0  & \alpha_{4,4}1 &\alpha_{4,4}4 &\alpha_{4,4}7\\[0.5em]
 0& 0&0 &0 &0  & \color{red}\alpha_{4,5}1 &\color{red}\alpha_{4,5}1 &\color{red}\alpha_{4,5}1\\[0.5em]
 \alpha_{1,6}8 & 0 &0& 0 &0  &\color{red}\alpha_{4,6}5 &\color{red}\alpha_{4,6}6 &\color{red}\alpha_{4,6}1\\[0.5em]
\alpha_{1,7}4& \alpha_{2,7}7 & \alpha_{2,7}8&0 &0 &\color{red}\alpha_{4,7}2 &\color{red}\alpha_{4,7}3 &\color{red}\alpha_{4,7}1 \\[0.5em]
\color{red}{\alpha_{1,8}1}& 0 &0&\alpha_{3,8}7 & \alpha_{3,8}8 &0&0& 0 \\[0.5em]
  \end{bmatrix}
 \end{matrix} .%\label{ex-90}
\end{equation}
Note that  the red sub-matrix in each $\boldsymbol{\Psi}^{(\pi_k^{*})}_k,\: k \in [4]$ is nonsingular.
\end{example}

\subsubsection{Assignment of $\alpha_{k,n_{k,j}}$ Variables by the Master Node} Now we prove that there exist $\alpha_{k,n_{k,i}}, k \in[K],\:n_{k,i}\in \A_k$ parameters such that the matrix $\mathbf{A}$,
used in the set of linear equations \eqref{4-7},  is nonsingular, such that
matrix $\mathbf{V}^{(\pi^*)}$ is nonsingular.  In this proof, we rely on the following lemma.

\begin{lemma}
\label{lem:full rank}
Let $\mathbf{V}$ be a square matrix that can be written as
\begin{gather}
\mathbf{V}= \diag(\pmb{\zeta}) \mathbf{C}+\mathbf{D},
\label{eq:6_63}
\end{gather}
where $\pmb{\zeta} \in \F^{n\times1}$,  $\mathbf{C}, \mathbf{D} \in \F^{n\times n}$, for some integer $n$,  and $\mathbf{C}$ is a full rank matrix. In addition,  $\mathbf{D}$ is not a function of $\pmb{\zeta}$. Then,  there exists a vector  $\pmb{\zeta}$ such that
 $\mathbf{V}$ is a full rank matrix.
\end{lemma}
The proof of Lemma~\ref{lem:full rank} can be found in Appendix \ref{A_3}. \\
We now aim to prove $\mathbf{V}^{(\pi^*)}$  satisfies the conditions of Lemma~\ref{lem:full rank}.  We define matrix $\hat{\mathbf{C}}$ as follows:
\begin{align}
\hat{\mathbf{C}} [n,\ell]=
\mathbf{V}^{(\pi^*)} [n,\ell], \label{eq:6_63_1}
\end{align}
if $n \in  \mathcal{Z}^{(\pi_k^{*})}_k$ and  $\sum_{i=1}^{k-1}R'_i+1\leq  \ell \leq \sum_{i=1}^{k}R'_i]$, for some $k \in [K]$. Otherwise
$\hat{\mathbf{C}} [n,\ell] =0$. Thus,
\begin{align}
\hat{\mathbf{C}} [n,\ell]=
\left\{
\begin{matrix}
\mathbf{V}^{(\pi^*)} [n,\ell] & n \in  \mathcal{Z}^{(\pi^{*}_k)}_k,  \sum_{i=1}^{k-1}R'_i+1\leq  \ell \leq \sum_{i=1}^{k}R'_i]\\
0 & \textrm{otherwise}
\end{matrix}.
\right.\label{eq:6_63_2}
\end{align}
We also define
\begin{equation}
\mathbf{D} = \mathbf{V}^{(\pi^*)}  - \hat{\mathbf{C}}.\label{eq:6_63_20}
\end{equation}
Then, regarding \eqref{eq:6_63_2}, we observe that
\begin{align}
\hat{\mathbf{C}} (\Z_k^*, \sum_{i=1}^{k-1}R'_i+1:\sum_{i=1}^{k}R'_i)   =   \mathbf{\Psi}^{(\pi^{*}_k)}_{k} (\mathcal{Z}^{(\pi^{*}_k)}_k, [R'_k])   \ \ k \in [K].\label{eq:6_63_4}
\end{align}
Recall that from \eqref{coherent-1}, \eqref{coherent-2} and \eqref{coherent-3} for any $k, k' \in [K]$, $k \neq k'$, $\Z_k^* \cap \Z_{k'}^* =\emptyset$, and also $\{\sum_{i=1}^{k-1}R'_i+1:\sum_{i=1}^{k}R'_i \} \cap \{\sum_{i=1}^{k'-1}R'_i+1:\sum_{i=1}^{k'}R'_i \}  =\emptyset$, we observe that $\hat{\mathbf{C}}$ includes $K$ nonzero blocks, as stated above, where these blocks do not overlap in their corresponding rows or columns in $\hat{\mathbf{C}}$. Moreover, due to \eqref{coherent-4}, these blocks are full rank.  In addition, regarding  \eqref{eq:5_12_12_1} we have $\mathbf{\Psi}^{(\pi*_k)}_{k} (\mathcal{Z}^{(\pi^*_k)}_k, [R'_k])  =    \diag(\boldsymbol{\alpha}_k( \Z_k^*))    \mathbf{V}^{(\pi^*_k)}_{k} (\mathcal{Z}^{(\pi_k^*)}_k,  [R'_k])$,  $k \in [K]$. Thus,
\begin{align}
\hat{\mathbf{C}} [\Z_k^*, \sum_{i=1}^{k-1}R'_i+1:\sum_{i=1}^{k}R'_i]    =    \diag(\boldsymbol{\alpha}_k( \Z_k^*))    \mathbf{V}^{(\pi^*_k)}_{k} (\mathcal{Z}^{(\pi_k^*)}_k,  [R'_k])  \ \ k \in [K].\label{eq:6_63_5}
\end{align}
We define vector $\pmb{\zeta}$, as $\zeta_n=   \alpha_{k,n}$, if $n \in \Z_k^*$ for some $k  \in [K]$, otherwise   $\zeta_n=0$.  Then, we have,
\begin{align}
\hat{\mathbf{C}} =    \diag(\pmb{\zeta})   \mathbf{C},\label{eq:6_63_7}
\end{align}
where
\begin{align}
\mathbf{C} [\Z_k^*, \sum_{i=1}^{k-1}R'_i+1:\sum_{i=1}^{k}R'_i]    =    \mathbf{V}^{(\pi^*_k)}_{k} (\mathcal{Z}^{(\pi_k^*)}_k,  [R'_k])  \ \ k \in [K], \label{eq:6_63_9}
\end{align}
and other entries of $\mathbf{C}$ are all zero.  In addition, $\mathbf{C}$ is nonsingular. Moreover, entries of $\pmb{\zeta}$ do not appear in $\mathbf{D}$.
 Thus, according to Lemma~\ref{lem:full rank}, there exists $\pmb{\zeta} \in  \mathbb{F}^N$, such that  $\mathbf{V}^{(\pi^*)} =  \diag(\pmb{\zeta}) \mathbf{C}+\mathbf{D}$ is nonsingular.

\begin{example}
Let us decompose $\mathbf{V}^{(\pi*)}$. According to \eqref{eq:6_63_2}, \eqref{eq:6_63_20} and \eqref{eq:6_63_7}, we need following matrices:
\begin{equation}
\begin{matrix}
\mathbf{C}
 =
 \begin{bmatrix}
 0 & 0 & 0 & { \color{red} 1}  & {\color{red} 1} & 0&0 &0\\[0.5em]

%1
 0 &0 &0& \color{red} 8 & \color{red}7&  0  &0 & 0  \\[0.5em]
%
%2
 0 & \color{red}8 & \color{red}7 &  0& 0&0 & 0 &0 \\[0.5em]
%3
 0 & \color{red}1 & \color{red}1 & 0 & 0  & 0 & 0 &0\\[0.5em]
 0& 0&0 &0 &0  & \color{red}1 &\color{red} 1 &\color{red}1\\[0.5em]
 0 & 0 &0& 0 &0  &\color{red}5 &\color{red}6 &\color{red}1\\[0.5em]
0& 0 &  0 &0 &0 &\color{red}2 &\color{red}3 &\color{red}1 \\[0.5em]
\color{red}{1}& 0 &0&0 & 0 &0&0& 0 \\[0.5em]
  \end{bmatrix}
 \end{matrix},
 %\\\label{ex-90}
\end{equation}
\begin{equation}
    \boldsymbol{\zeta} =
\begin{bmatrix}
\alpha_{3,1} &  \alpha_{3,2} &  \alpha_{2,3} & \alpha_{2,4}  & \alpha_{4,5} & \alpha_{4,6} & \alpha_{4,7} & \alpha_{1,8}
\end{bmatrix},
\end{equation}
\begin{equation}
    \begin{matrix}
\mathbf{D}^{}
 =
 \begin{bmatrix}
  \alpha_{1,1}7 &  \alpha_{2,1}4 &\alpha_{2,1} 0& 0  &0 & \:\:\:\:\:\:0\:\:\:\:\:\:\:\:\:\:\:\: & \:\:\:\:\:\:0\:\:\:\:\:\:\:\:\: & \:\:\:\:\:\:\:\:\:\:\:\:0\:\:\:\:\:\:\:\:\:\\[0.5em]

%1
 0 &0 &0& 0 & 0&  \alpha_{4,2}6 &\alpha_{4,2}4 &\alpha_{4,2}10 \\[0.5em]
%
%2
 0 & 0 & 0 & \alpha_{3,3}4& \alpha_{3,3}0&0 & 0 &0^{
} \\[0.5em]
%3
 0 & 0 & 0 & 0 & 0  & \alpha_{4,4}1 &\alpha_{4,4}4 &\alpha_{4,4}7\\[0.5em]
 0& 0&0 &0 &0  &0 &0 &0\\[0.5em]
 \alpha_{1,6}8 & 0 &0& 0 &0  &0 &0 &0\\[0.5em]
\alpha_{1,7}4& \alpha_{2,7}7 & \alpha_{2,7}8&0 &0 &0&0&0 \\[0.5em]
0& 0 &0&\alpha_{3,8}7 & \alpha_{3,8}8 &0&0& 0 \\[0.5em]
  \end{bmatrix}
 \end{matrix}.
 %\\\label{ex-90}
\end{equation}
As can be seen $\pmb{\zeta}$ entries are independent from  $\mathbf{D}$ entries hence regarding  Lemma~\ref{lem:full rank}, there exists $\pmb{\zeta} \in  \mathbb{F}^N$, such that  $\mathbf{V}^{(\pi^*)} =  \diag(\pmb{\zeta}) \mathbf{C}+\mathbf{D}$ in our example becomes nonsingular.
\end{example}

\subsection{Constructing the Encoding and Decoding Functions}
\label{Encoding_Decoding_1}
In this subsection, we show that how the encoding and decoding functions defined in parts  \ref{Encoding} and \ref{Decoding} are applied.
\begin{example}\emph{Encoding Function}:
Referring to  Example Part \ref{initilization, encoding}, now we want to construct the encoding equation which must be solved by the master node. The encoding equation is as follows,
 \begin{align}
     \mathbf{A}\mathbf{b}=\mathbf{s},
\label{ex-110}
 \end{align}
 and  $b$ is  unknown  and consists of,
\begin{align}
    \mathbf{b}&=[ P_{1,1}, P_{1,2}, P_{1,3},P_{2,2},P_{2,3},P_{2,2},P_{2,3},P_{4,3},P_{4,4},\mathbf{y}^\intercal]^\intercal.
    \label{ex-111}
\end{align}
Having $\gamma_{k,i},\:k \in [4],\:i \in \A_k$ and $\alpha_k,\: k \in [4]$ the known vector $\mathbf{s}$ becomes,
\begin{align}
    \mathbf{s} &=  [\mathbf{s}_1^\intercal,\mathbf{s}_2^\intercal,\mathbf{s}_3^\intercal,\mathbf{s}_4^\intercal]^ \intercal,\label{ex-112}\\
    \mathbf{s}_1 &= -[1,1,1,1]^\intercal\:,\:%\label{ex-62}\\
    \mathbf{s}_2 =- [5,1,6,4]^\intercal\:,\:%\label{ex-63}\\
     \mathbf{s}_3 = -[4,4,4,4]^\intercal\:,\:%\label{ex-64-1}\\
      \mathbf{s}_4 = -[3, 5, 7,
    10,10]^\intercal,\label{ex-113}\\
      \mathbf{A}&=
\left[ \begin{array}{cccc|c}
  \mathbf{B}^{(\pi^{*}_1)\:\intercal}_1& & & &\mathbf{\Upsilon}_1\\
   & \mathbf{B}^{(\pi^{*}_2)\:\intercal}_2& & &\mathbf{\Upsilon}_2\\
  & & \mathbf{B}^{(\pi^{*}_3)\:\intercal}_3& &\mathbf{\Upsilon}_3\\
 & & & \mathbf{B}^{(\pi^{*}_4)\:\intercal}_4 &\mathbf{\Upsilon}_4\\
\end{array}\right].\label{ex-114}
\end{align}
 Also we have,
\begin{align}
\mathbf{B}^{(\pi^{*}_1)}_1 &=
\begin{bmatrix}
8^{4 } & 8^{2 } & 8^{3 }& 8^{1}
\\
9^{4 } & 9^{2 } & 9^{3 }& 9^{1}
\\
6^{4 } & 6^{2 } & 6^{3 }& 6^{1}
\end{bmatrix}\:,\:%\label{ex-115}\\
\mathbf{B}^{(\pi^{*}_2)}_2 =
\begin{bmatrix}
9^{3 } & 9^{2 } & 9^{1 }& 9^{4}
\\
6^{3 } & 6^{2 } & 6^{1 }& 6^{4}
\end{bmatrix},
\label{ex-116}\\
\mathbf{B}^{(\pi^{*}_3)}_3 &=
\begin{bmatrix}
9^{1} & 9^{2 } & 9^{3 }& 9^{4}
\\
6^{1} & 6^{2 } & 6^{3 }& 6^{4}
\end{bmatrix}\:,\:%\label{ex-117}\\
\mathbf{B}^{(\pi^{*}_4)}_4 =
\begin{bmatrix}
6^{4} & 6^{5 } & 6^{1 }& 6^{2} & 6^{3}
\\
4^{4} & 4^{5 } & 4^{1 }& 4^{2} & 4^{3}
\end{bmatrix},
\label{ex-118}\\
\boldsymbol{\Upsilon}_1 &=
\begin{bmatrix}
1& 0 & 0 &0& 0 &  0 &0& 0\\
0& 0 & 0 &0& 0 & 1 &0& 0\\
0& 0 & 0 &0& 0 & 0 &1& 0\\
0& 0 & 0 &0& 0 & 0 &0& 1\\
\end{bmatrix}\:,\:%\label{ex-119}\\
\boldsymbol{\Upsilon}_2 =
\begin{bmatrix}
1& 0 & 0 &0& 0 &  0 &0& 0\\
0& 0 & 7&0& 0 & 0 &0& 0\\
0& 0 & 0 &1& 0 & 0 &0& 0\\
0& 0 & 0 &0& 0 & 0 &8& 0\\
\end{bmatrix},\label{ex-120}\\
\boldsymbol{\Upsilon}_3 &=
\begin{bmatrix}
1& 0 & 0 &0& 0 &  0 &0& 0\\
0& 1 & 0&0& 0 & 0 &0& 0\\
0& 0 &1 &0& 0 & 0 &0& 0\\
0& 0 & 0 &0& 0 & 0 &0& 1\\
\end{bmatrix}\:,\:%\label{ex-121}\\
\boldsymbol{\Upsilon}_4 =
\begin{bmatrix}
0& 1 & 0 &0& 0 &  0 &0& 0\\
0& 0 & 0& 1& 0 & 0 &0& 0\\
0& 0 &0 &0& 1 & 0 &0& 0\\
0& 0 & 0 &0& 0 & 1 &0& 0\\
0& 0 & 0 &0& 0 & 0 &1& 0
\end{bmatrix}.\label{ex-122}
\end{align}

Now, by solving \eqref{ex-110}, we obtain,
\begin{equation}
    \mathbf{b}=[5,2,7,2,4,9,7,6,5,5,5,8,7,3,2,2,9]^\intercal.
\end{equation}
Therefore, the encoded (stored) data are,
\begin{equation}
{Y}_1=5,\: {Y}_2=5,\: {Y}_3=8,\: {Y}_4=7,\: {Y}_5=3,\: {Y}_6=2,\: {Y}_7=2,\: {Y}_8=9.
\end{equation}
Note that $\phi^{M},\:M=1$, the encoding function maps $W_{[4]}$ to $Y_{[8]}$.
\end{example}
Now notice how each user applies its decoding function in the following  Example Part.
\begin{example}
   First, we rewrite \eqref{4-3-1-4} as,
   \begin{align}
       \mathbf{D}^{(\pi^{*}_1)}_1 \mathbf{b}_1 = \mathbf{y}_1, \label{eq:ex-dec}
   \end{align}
   where,
   \begin{align*}
       \mathbf{D}^{(\pi^{*}_1)}_1 &= \begin{bmatrix}
1 &(8^4)^{1 } & (8^4)^{2 } & (8^4)^{3 }
\\
1 &(8^2)^{1 } & (8^2)^{2 } & (8^2)^{3 }\\
1 &(8^3)^{1 } & (8^3)^{2 } & (8^3)^{3 }\\
1 &(8)^{1 } & (8)^{2 } & (8)^{3 }\\
\end{bmatrix}\;,\;%\label{ex-80}\\
\mathbf{b}_1=[\hat{W}_{1,0}, \hat{P}_{1,1},\hat{P}_{1,2},\hat{P}_{1,3}]^\intercal,\\
\mathbf{y}_1&=[-\alpha_{1,1}Y_1,-\alpha_{1,6}Y_6,-\alpha_{1,7}Y_7,-\alpha_{1,8}Y_8]^\intercal = [6,9,9,2]^\intercal.
   \end{align*}
Solving \eqref{eq:ex-dec}, we obtain,
\begin{equation}
    \mathbf{b}_1 =[1,5,2,7]^{\intercal}.
\end{equation}
Therefore, $\hat{W}_{1,0}=1$, which is identical to  ${W}_{1,0}$. Note that the decoding function $\theta^{1}_{1}$ maps $Y_{\A_1}=\{y_1,y_6,y_7,y_8\}$ to $\hat{W}_{1}$. Similarly for User $2$ we have,
\begin{align}
   \mathbf{D}^{(\pi^{*}_2)}_2 \mathbf{b}_2 = \mathbf{y}_2,
\end{align}
   where
   \begin{align}
       \mathbf{D}^{(\pi^{*}_2)}_2 &= \begin{bmatrix}
1 &(8^3)^{1 } & (8^3)^{2 } & (8^3)^{3 }
\\
1 &(8^2)^{1 } & (8^2)^{2 } & (8^2)^{3 }\\
1 &(8^1)^{1 } & (8^1)^{2 } & (8^1)^{3 }\\
1 &(8^4)^{1 } & (8^4)^{2 } & (8^4)^{3 }\\
\end{bmatrix}\;,\;%\label{ex-80}\\
\mathbf{b}_2=[\hat{W}_{2,0},\hat{W}_{2,1} \hat{P}_{2,2},\hat{P}_{2,3}]^\intercal,\\
\mathbf{y}_2&=[-\alpha_{2,1}Y_1,-\alpha_{2,3}Y_3,-\alpha_{2,4}Y_4,-\alpha_{2,7}Y_7]^\intercal = [6,10,4,6]^\intercal.
   \end{align}
This results in $\mathbf{b}_2 =[2,6,2,4]^{\intercal}$.
For User 3 we have,
    \begin{align}
       \mathbf{D}^{(\pi^{*}_3)}_3 \mathbf{b}_3 = \mathbf{y}_3,
   \end{align}
   where
   \begin{align}
       \mathbf{D}^{(\pi^{*}_3)}_3 &= \begin{bmatrix}
1 &(8^1)^{1 } & (8^1)^{2 } & (8^1)^{3 }
\\
1 &(8^2)^{1 } & (8^2)^{2 } & (8^2)^{3 }\\
1 &(8^3)^{1 } & (8^3)^{2 } & (8^3)^{3 }\\
1 &(8^4)^{1 } & (8^4)^{2 } & (8^4)^{3 }\\
\end{bmatrix}\;,\;%\label{ex-80}\\
\mathbf{b}_3=[\hat{W}_{3,0},\hat{W}_{3,1} \hat{P}_{3,2},\hat{P}_{3,3}]^\intercal,\\
\mathbf{y}_3&=[-\alpha_{3,1}Y_1,-\alpha_{3,1}Y_2,-\alpha_{3,3}Y_3,-\alpha_{3,8}Y_8]^\intercal = [6,6,3,2]^\intercal,
   \end{align}
which leads to $\mathbf{b}_3 =[4,0,9,7]^{\intercal}$.
%Solving the equations we have,
%   \begin{equation}
%        \mathbf{b}_3 =[4,0,9,7]^{\intercal}.
%   \end{equation}
Finally, for User $4$ we assert
    \begin{align}
       \mathbf{D}^{(\pi^{*}_4)}_4 \mathbf{b}_4 = \mathbf{y}_4,
   \end{align}
   where
   \begin{align}
       \mathbf{D}^{(\pi^{*}_4)}_4 &= \begin{bmatrix}
1 &(8^4)^{1 } & (8^4)^{2 } & (8^4)^{3 }  & (8^4)^{4 }
\\
1 &(8^5)^{1 } & (8^5)^{2 } & (8^5)^{3 }  & (8^5)^{4 }\\
1 &(8^1)^{1 } & {(8^1)}^{2 } & (8^1)^{3 }  & (8^1)^{4 }\\
1 &(8^2)^{1 } & (8^2)^{2 } & (8^2)^{3 }  & (8^2)^{4}\\
1 &(8^3)^{1 } & (8^3)^{2 } & (8^3)^{3 }  & (8^3)^{4 }\\
\end{bmatrix}\;,\;%\label{ex-80}\\
\mathbf{b}_4=[\hat{W}_{4,0},\hat{W}_{4,1},\hat{W}_{4,2} \hat{P}_{4,3},\hat{P}_{4,4}]^\intercal,\\
\mathbf{y}_4&=[-\alpha_{4,2}Y_2,-\alpha_{4,4}Y_4,-\alpha_{4,5}Y_5,-\alpha_{4,6}Y_6,-\alpha_{4,7}Y_7]^\intercal = [6,6,3,2]^\intercal,
   \end{align}
which has the solution as $\mathbf{b}_4 =[3,7,5,6,5]^{\intercal}$.
\end{example}

\subsection{Proof of Correctness}
\label{Correctness_1}
Note that User $k$ has access to ${|\A_k|}$ point of $g_{\mathbf{w}_k,\:\mathbf{p}_k}$ polynomial with degree  ${|\A_k|-1}$, which each of them consists of $|\A_k|$ elements of $\mathbf{w}_k,\:\mathbf{p}_k$. In the retrieval phase, $\mathbf{w}_k$ is the desired secret message of the user. As can be seen from \eqref{4-3-1} and \eqref{4-3-1-4}, both encoding and decoding system of equations imply the same relation between $\mathbf{w}_k,\mathbf{p}_k$, $\hat{\mathbf{{w}}}_k,\hat{\mathbf{{p}}_k}$ and stored data $\mathbf{y}$. Since the decoding equation, formed by user $k$, has a unique solution due to the nonsingularity of  Vandermonde matrix  and also due to Subsection \ref{Feasibility_1} it is known that there exists a feasible scheme to uniquely encode the stored data, namely  $\mathbf{y}$, Consequently, we can assert that $\hat{\mathbf{{w}}}_k=\mathbf{w}_k$ and the correctness condition holds.
\label{Correctness}

\subsection{Proof of Privacy}
\label{Privacy_1}
In this subsection, we prove that the achievable scheme satisfies the privacy condition in~\eqref{2-7}.
Let us begin by the following lemma,
\begin{lemma}\label{lemma5}
Symbols $Y_1,Y_2,\hdots,Y_N$, stored in the storage nodes 1 to $N$,  are mutually independent and full entropy. In other words:
\begin{align}
\label{l2}
 H(Y_1,Y_2,\hdots,Y_N)& = N\log (|\mathbb{F}|).
\end{align}
\end{lemma}
\begin{IEEEproof}
Let us define
\begin{align}
 \mathcal{P} \triangleq \bigcup_{k \in [K]} \{P_{k,R_k},\hdots,P_{k,|\A_k|-1}\}.
\end{align}

\begin{align}
 \mathcal{O} \triangleq \bigcup_{k \in [K]} \{P_{k,R_k},\hdots,P_{k,R'_k-1}\}.
 \end{align}
Note that $R'_k \leq |\A_k|$, $k \in [K]$, thus  $\mathcal{O} \subset  \mathcal{P}$.
Then, we have
%Note that by invoking $\phi^{M}$ on secret messages $W_1,W_2,\hdots,W_K$ and randomly chosen $\mathcal{O} \triangleq \{P_{k,R_k},\hdots,P_{k,R'_k-1}\}$, The stored data namely, $Y_1,Y_2,Y_3,\hdots,Y_N$ are uniquely determined. Also by invoking $\theta^{M}_{k}$ for all $k\in[K]$ a unique solution for the secret messages could be determined. Therefore there exist a one-to-one mapping between secret messages plus $\mathcal{O}$ and $Y_{1}, Y_{2}, \hdots, Y_{N}$. From mathematical perspective we have,
\begin{align}
H(Y_1,Y_2,\hdots,Y_{N})\overset{(a)}=& H(Y_1,Y_2,\hdots,Y_{N})+ H(W_1,W_2,\hdots,W_{K},\mathcal{P}|Y_1,Y_2,\hdots,Y_{N})  \label{4-20-1}\\
 \overset{(b)}\geq& H(Y_1,Y_2,\hdots,Y_{N})+ H(W_1,W_2,\hdots,W_{K},\mathcal{O}|Y_1,Y_2,\hdots,Y_{N})  \label{4-20-1}\\
  = &H(W_1,W_2,\hdots,W_{K},Y_1,Y_2,\hdots,Y_{N},\mathcal{O})\\
   \geq  & H(W_1,W_2,\hdots,W_{K},\mathcal{O})\\
  \overset{(c)} = & H(W_1,W_2,\hdots,W_{K})+ H(\mathcal{O})\\
  \overset{(d)}= & (\sum^K_{k=1}R'_k) \log (|\mathbb{F}|)\overset{(e)}= N\log (|\mathbb{F}|),\label{4-22-1}
\end{align}
where (a) follows from the fact that having access to $Y_1,Y_2,\hdots,Y_{N}$, we can recover $W_1,W_2,\hdots,W_{K}$ and  $\mathcal{P} \triangleq \bigcup_{k \in [K]} \{P_{k,R_k},\hdots,P_{k,|\A_k|_k-1}\}$ as proven in the correctness of the decoding phase of the proposed scheme  in Subsection \ref{Decoding}; (b) is true due to $\mathcal{O} \subset  \mathcal{P}$; (c) holds since as explained in Subsection~\ref{Feasibility_1}, $W_1,W_2,\hdots,W_{K}$ and variables in $\mathcal{O}$ are chosen independently; (d) holds because $W_1,W_2,\hdots,W_{K}$ and variables in $\mathcal{O}$ are chosen independently and  uniformly at random from $\mathbb{F}$; and (e) holds since according to \eqref{4-22-222}, we have $\sum^{K}_{i=1} R'_i=N$.

On the other hand, we know  $H(Y_1,Y_2,\hdots,Y_{N}) \leq N \log (|\mathbb{F}|)$, thus $H(Y_1,Y_2,\hdots,Y_{N}) = N \log (|\mathbb{F}|)$.
\end{IEEEproof}

Now consider any two Users $k$ and $k'$ with access sets $\mathcal{A}_k$ and $\mathcal{A}_{k'}$, respectively. We prove that $H(W_{{k}}|Y_{\mathcal{A}_{k'}}) = H(W_{{k}})= R_k \log(|\mathbb{F}|)$, meaning that User $k'$ learn nothing about the secret message of User $k$.

First note that $R_k \leq |{\A_k} \backslash \A_{k'}|$.  Let $\mathcal{A}_k^{(-k')}$ be an arbitrary subset of ${\A_k} \backslash \A_{k'}$ with size $R_K$.  In other words, we have  $\mathcal{A}_k^{(-k')} \subset {\A_k}  \backslash \A_{k'}$, and $|\mathcal{A}_k^{(-k')}|=R_k$.
Then, we have
\begin{align}
   H(W_{{k}}|Y_{\mathcal{A}_{k'}})\overset{(a)}{\geq} & H( W_{{k}}|Y_{[N] \backslash \mathcal{A}_k^{(-k')} }) \\
    \overset{(b)}{=} &H(W_{{k}}|Y_{[N] \backslash \mathcal{A}_k^{(-k')} } ) +    H(\mathbf{p}_k|Y_{[N] \backslash \mathcal{A}_k^{(-k')} } ,W_k)+   H(Y^{(-k')}_{k}|W_{{k}},Y_{[N] \backslash \mathcal{A}_k^{(-k')} },\mathbf{p}_k)\\
     {=} & H(W_k,Y_{ \mathcal{A}_k^{(-k')} } ,\mathbf{p}_k|Y_{[N] \backslash \mathcal{A}_k^{(-k')} })\\
     {=} & H(Y_{ \mathcal{A}_k^{(-k')} }|Y_{[N] \backslash \mathcal{A}_k^{(-k')} })+ H(W_k,\mathbf{p}_k|Y_{[N]})\\
   \geq & H(Y_{ \mathcal{A}_k^{(-k')} }|Y_{[N] \backslash \mathcal{A}_k^{(-k')} })\overset{(c)}{=} H(Y_{ \mathcal{A}_k^{(-k')} })\overset{(d)}{ =} R_k \log(|\mathbb{F}|),\label{13}
\end{align}
where (a) holds since $Y_{\mathcal{A}_{k'}} \subset Y_{[N] \backslash \mathcal{A}_k^{(-k')} }$, (c) and (d) follow from Lemma  \ref{lemma5}. To establish (b), we argue that
    \begin{align}
    &   H(\mathbf{p}_k|Y_{[N] \backslash \mathcal{A}_k^{(-k')} },W_k)=0,\label{4-3-33333}\\
  &     H(Y^{(-k')}_{k}|W_{{k}},Y_{[N] \backslash \mathcal{A}_k^{(-k')} },\mathbf{p}_k)=0\label{4-3-333}.
       \end{align}
The reason is that from \eqref{4-3-1-1-1}, and having $W_{k},Y_{[N] \backslash \mathcal{A}_k^{(-k')} }$, we can form the following set of linear equations,
 \begin{align}
   P_{k,R_k+1}\gamma_{k,i}^{R_k}+P_{k,R_k+2}\gamma_{k,i}^{R_k+1}+\hdots+P_{k,{|\A_k|}}\gamma_{k,i}^{{|\A_k|}-1}&= -(g_{\mathbf{w}_k}(\gamma_{k,i})+\alpha_{k,n_{k,i}}Y_{n_{k,i}}),\:  \forall n_{k,i} \in \A_k  \backslash \mathcal{A}_k^{(-k')},
   \label{4-3-44}
  \end{align}
  where the right-hand side is known. In the above equations, we have $|\A_k|-R_k$ variables and $|\A_k  \backslash \mathcal{A}_k^{(-k')}| = |\A_k|-R_k$ equations.   Since the coefficients of \eqref{4-3-44} form a Vandermonde full-rank matrix, $\mathbf{p}_k$ can be solved for, and thus \eqref{4-3-33333} holds. On the other hand,  having  $\mathbf{p}_k$ and $W_k$, we can calculate $Y_{ \mathcal{A}_k}$ and thus $Y_{ \mathcal{A}_k^{(-k')} }$. As a result, \eqref{4-3-333} holds.

% For proof of \eqref{4-3-33333}, the same argument as of Section \ref{Theorem 1} about  solvability of \eqref{4-3-1-4} holds. Having $W_{{k}},Y_{[N]}\backslash Y^{(-k')}_{k}$ determined will make the right side of \eqref{4-3-44} known and $\mathbf{p}_k$ are unknown variables. Since we know only $Y_{[N]}\backslash Y^{(-k')}_{k}$ variables, we  can construct $|\mathcal{A}_i|-R_i$ of the right-hand side of these linearly independent equations and because the number of unknown variables is also $|\A_k|-R_k$, it is sufficient for evaluation of $\mathbf{p}_K$. Note that we already know these equations are compatible because of achievability scheme. Therefore $\eqref{4-3-33333}$ results.

\subsection{Memory Sharing}
\label{Memory Sharing}
\begin{lemma}
Consider two achievable rate tuples $(R_1,R_2,\hdots,R_K)$ and $(\hat{R}_1,\hat{R}_2,\hdots,\hat{R}_K)$. Then for any $ \omega \in [0, 1]$, the rate tuple $\omega (R_1,R_2,\hdots,R_K) + (1-\omega) (\hat{R}_1,\hat{R}_2,\hdots,\hat{R}_K)$ is achievable.
\end{lemma}
 \begin{IEEEproof}
Assume that the size of the memory is $M$. We split each memory unit into two parts with size  $\omega M$ and $ (1-\omega) M$. We then apply the first achievable scheme to store secrets with sizes $(R_1,R_2,\hdots,R_K) \omega M$ in the first part. Similarly, we use the second scheme to store the secrets with sizes $(\hat{R}_1,\hat{R}_2',\hdots,\hat{R}_K') (1-\omega)M$. Therefore, the size of the secrets stored are equal to $(R_1,R_2,\hdots,R_K) \omega M+ (\hat{R}_1,\hat{R}_2,\hdots,\hat{R}_K) (1-\omega)M$, and we achieve the rate of
$(R_1,R_2,\hdots,R_K) \omega + (\hat{R}_1,\hat{R}_2,\hdots,\hat{R}_K) (1-\omega)$.
\end{IEEEproof}

\section{The Converse Proof}\label{sec: Converse 1}
\begin{figure}
\centering
\includegraphics[scale=0.75]{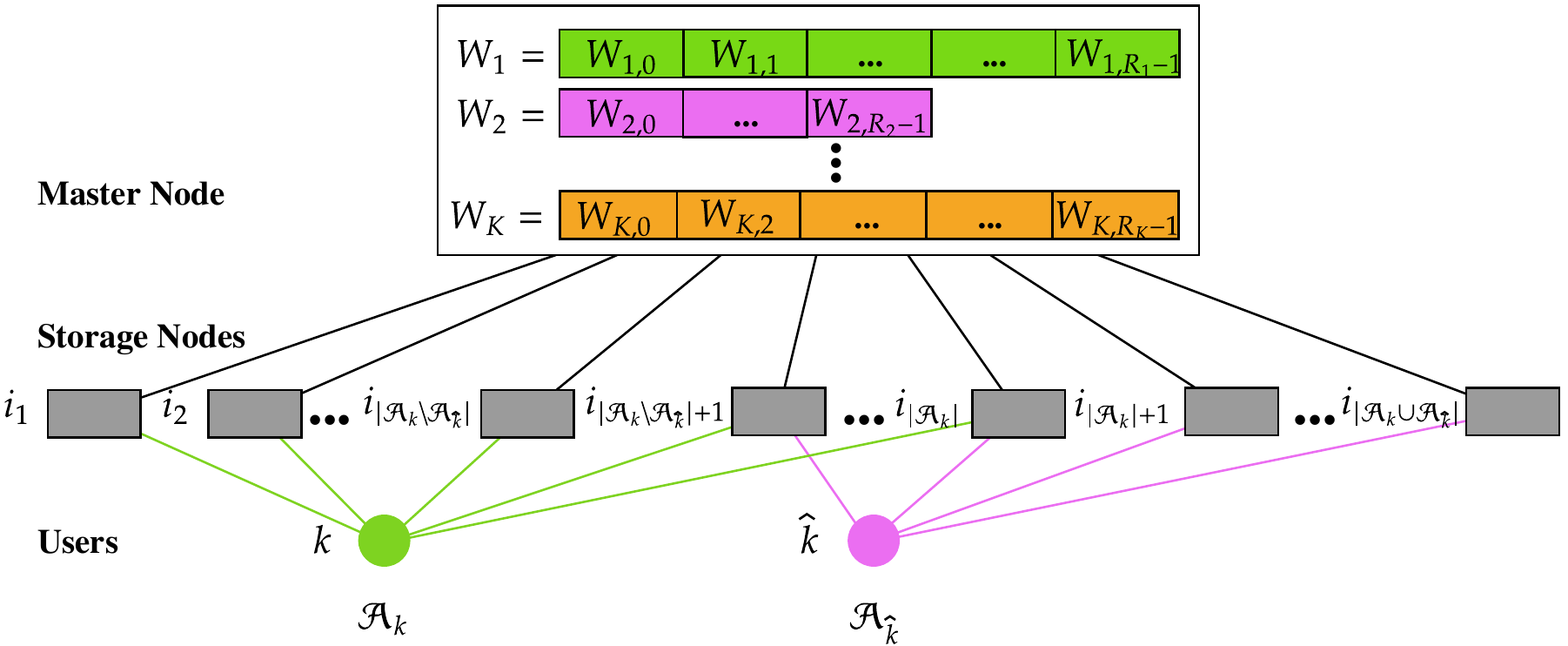}
\caption{Structure considered for the converse proof of Theorem~\ref{Theorem 1}.} \label{fig:3}
\end{figure}
Consider two Users $k$ and $\hat{k}$ as shown in Fig.~\ref{fig:3}. The master node intends to privately transmit $W_k$ with the size of $R_k$ $q$-array bits to User $k$. The other user, $\hat{k}$, acts as an eavesdropper and intends to obtain information about $W_k$. First, note that from the correctness and privacy conditions in \eqref{2-6-1} and \eqref{2-7}, respectively, we obtain:
\begin{align}
H(W_{k}|Y_{\A_{k}\cap\A_{k'}},Y_{\A_{k}\backslash \A_{k'}})&=M o(M),\label{4-33-4}\\
H(W_{k}|Y_{\A_{k}\cap\A_{k'}})&=H(W_{k}).\label{4-33-3}
\end{align}

Now, we have
\begin{align}
I(Y_{\A_k\backslash \A_{k'}};W_k|Y_{\A_{k}\cap\A_{k'}})&=%\label{4-33}
H(Y_{\A_k\backslash \A_{k'}}|Y_{\A_{k'}\cap\A_{k'}})-H(Y_{\A_{k}\backslash \A_{k'}}|W_{k},Y_{\A_{k}\cap\A_{k'}})\label{4-33}\\
&\overset{(a)}\leq H(Y_{\A_k\backslash \A_{k'}})%\label{4-33-55}\\
\overset{(b)}\leq M |\A_k\backslash\A_{k'}|\label{4-33-56},
\end{align}
where (a) is due to the fact that conditioning does not increase entropy, and (b) follows since the maximum entropy results when the stored data in $Y_{\A_k\backslash \A_{k'}}$ are independent and uniformly chosen.

Further, we have
\begin{align}
I(Y_{\A_k\backslash \A_{k'}};W_k|Y_{\A_{k}\cap\A_{k'}})&=H(W_{k}|Y_{\A_{k}\cap\A_{k'}})-H(W_{k}|Y_{\A_{k}\cap\A_{k'}},Y_{\A_{k}\backslash \A_{k'}}),\label{4-33-2}\\
&\overset{(a)}=H(W_{k}) - M o(M),\label{4-33-57}
\end{align}
where (a) follows from \eqref{4-33-4} and \eqref{4-33-3}. Combining \eqref{4-33-56} and \eqref{4-33-57} results in:
\begin{gather}
r_k - M o(M)\leq M|\A_k\backslash \A_{k'}|,\label{4-33-20}
\end{gather}
which in our problem, when $M$ tends to infinity, implies that,
\begin{equation}
  R_k \leq |A_k\backslash A_{k'}|.
   \label{4-34}
\end{equation}
This completes the converse proof for \eqref{3-1}.

%-----------------------------------------

Now, we prove \eqref{3-2}. Consider any arbitrary  set like $\mathcal{S} \in [K]$, where $\mathcal{S} =\{j_1, j_2,\hdots, j_{k}\}$. Note that from the problem setup and the correctness condition in \eqref{2-6-1}, we obtain:
\begin{align}
 H(W_{j_1},\hdots,W_{j_{k}}|Y_{\A_{j_1}},\hdots,Y_{\A_{j_{k}}}) &\leq |\S| M o(M), \label{4-40}\\
 H(W_{j_1},\hdots,W_{j_{k}})&=\sum_{j \in \S} r_j=M\sum_{j \in \S} R_j\label{4-40-1}
\end{align}

Now, we have
\begin{align}
I(Y_{\A_{j_1}},\hdots,Y_{\A_{j_{k}}}: W_{j_1},\hdots,W_{j_{k}})&= H(Y_{\A_{j_1}},\hdots,Y_{{\A_{j_{k}}}})-H(Y_{\A_{j_1}},\hdots,Y_{\A_{j_{k}}}|W_{j_1},\hdots,W_{j_{k}})\label{4-38}\\
&\leq M|\A_{j_1}\cup \hdots\cup \A_{j_{k}}|. \label{4-38-1}
\end{align}

Further, we have
\begin{align}
I(Y_{\A_{j_1}},\hdots,Y_{\A_{j_{k}}}: W_{j_1},\hdots,W_{j_{k}})&=H(W_{j_1},\hdots,W_{j_{k}})- H(W_{j_1},\hdots,W_{j_{k}}|Y_{\A_{j_1}},\hdots,Y_{{\A_{j_{k}}}})\label{4-37}\\
&\overset{(a)}\geq M\sum_{j \in \S} R_j - |\S| M o(M),\label{4-37-1}
\end{align}
where (a) follows from \eqref{4-40} and \eqref{4-40-1}.
Combining \eqref{4-33-56} and \eqref{4-33-57} results in:
\begin{equation}
    \sum_{j \in \S} R_j \leq  |\A_{j_1}\cup \hdots\cup \A_{j_{k}}|+|\S| o(M).
\end{equation}
Noting that $o(M)$ approaches zero as $M\rightarrow\infty$, this completes the proof.

\section{Conclusion}\label{sec: Conclusion}
In this paper, we considered the problem of distributed \emph{multi-user} secret sharing (DMUSS), consisting of a master node, $N$ storage nodes, and $K$ users. The master node, storing all secret messages, intends to convey each secret messages $W_k,k\in [K]$ to its corresponding User $k$ privately and correctly. We studied the general case, where each user has access to an arbitrary subset of storage nodes. We characterized the capacity region of the problem by deriving the optimal normalized size of secret messages that can be achieved.
Considering the DMUSS problem in presence of straggler or adversarial nodes would be an interesting future work.

\bibliographystyle{ieeetr}
\bibliography{ref}
\newpage
\appendices
\section{}\label{A_2}
\begin{figure}
\centering
\includegraphics[scale=0.75]{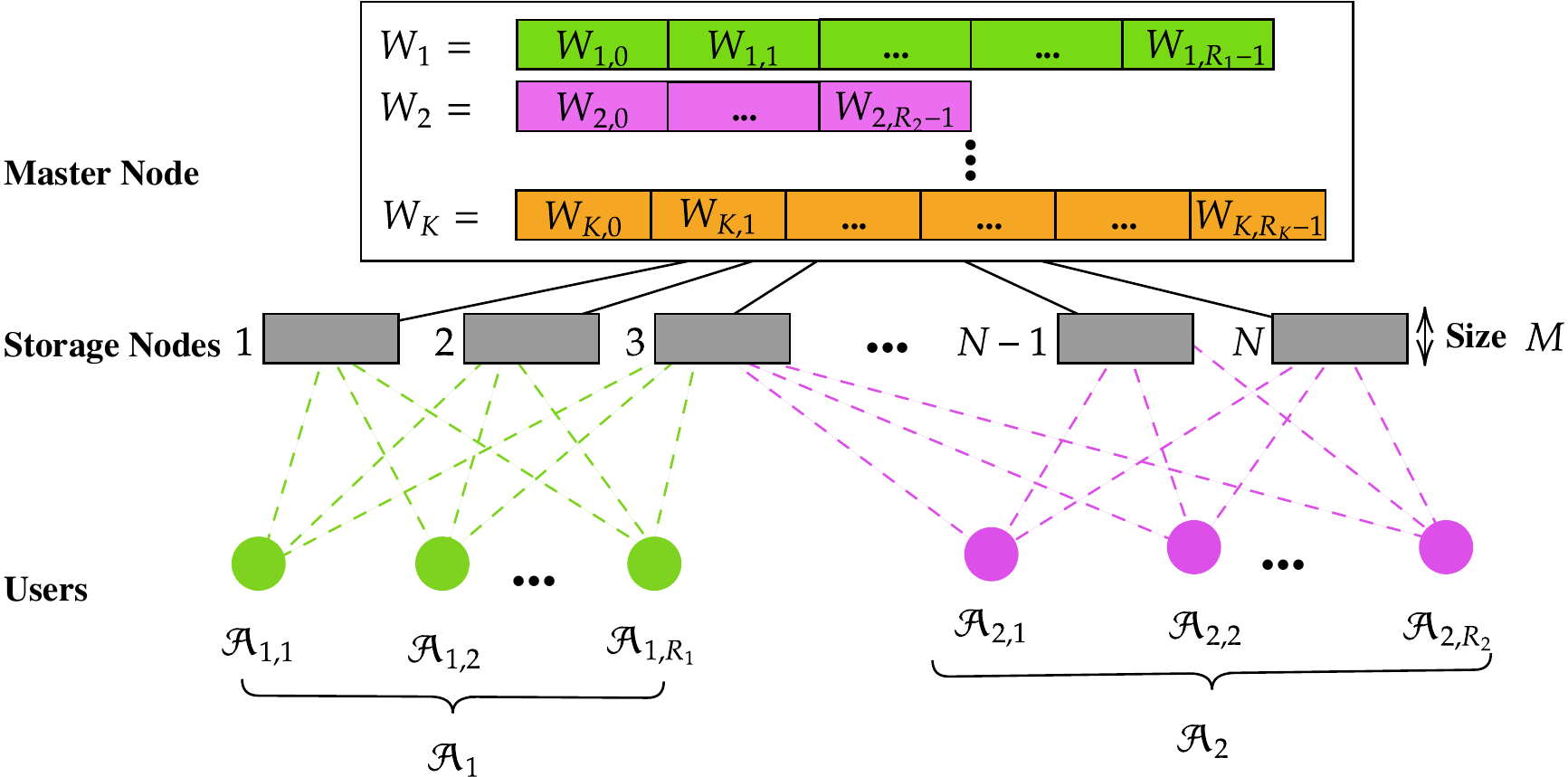}
\caption{Graphical Representation of Hall's Marriage Theorem.} \label{fig:Hall's}
\end{figure}
First, we clone User  $i \in [K]$'s node $R'_i$ times with the same access set (see Fig.~\ref{fig:Hall's}). To illustrate this idea
let us define
\begin{equation}
\A_{i,j}\triangleq \A_{i},\:    \forall i \in [K],\forall j \in [R'_i].\label{eq:5_20_1}
\end{equation}

Consider a family of sets $\A$ containing all of the $\A_{i,j},\: i \in [K],j \in [R'_i]$ sets. Note that members of $\A$ are counted with multiplicity.
A \emph{system of distinct representative} (SDR) for these sets is a $N$-tuple $(x_{1,1},x_{1,2},\hdots,x_{1,R'_1},\hdots,x_{K,1},x_{K,2},\hdots,x_{K,R'_K})$ of elements (where $N=\sum^{K}_{i=1}R'_i$) with the following properties:
\begin{enumerate}
\item \textbf{Representative Property}: $x_{i,j}\in \A_{i,j}$ for $i\in[K],\:j \in [R'_i]$.
\item \textbf{Distinct Property}: $x_{i_1,j_1} \neq  x_{i_2,j_2}$ for $(i_1,j_1)\neq (i_2,j_2)$.
\end{enumerate}
Also for any set of indices $\D \subseteq \{(i,j)|i \in [K],j \in [R'_i]\}$, we define
\begin{gather}
    \A(\D) \triangleq \cup_{d \in \D} \A_d.\label{eq:5_20_2}
\end{gather}
\begin{lemma}[Hall's Marriage Theorem \cite{r13}]
$\A$ as a family of finite sets has a system of distinct representative if and only if the following condition holds:
\begin{gather}
|\A(\D)|\geq|\D|,\: \forall \D \subseteq  \{(i,j)|i \in [K],j \in [R'_i]\}.\label{eq:5_20_3}
\end{gather}
 \end{lemma}
We claim that the condition in \eqref{eq:5_20_3} holds in our problem. For the proof of the claim, consider a subset $\D \subseteq  \{(i,j)|i \in [K],j \in [R_i]\}$. Define $\T \triangleq \{i|(i,j)\in  \D \}=\{i_1,i_2,\hdots, i_m\}$ and also  $l_e \triangleq |\{(i,j)\in \D|i_e \in \T \}|,\: \forall e \in [m]$ which shows the number of sets that their unification make $\A(\D)$. From \eqref{eq:5_20_1}, we conclude that
\begin{gather}
    \A(\D)= \A_{i_1}\cup\A_{i_2} \cup \hdots \cup \A_{i_m}\label{eq:5_20_5},
\end{gather}
% its trivial that the union of the same sets is equal to all of them.
Also from the definition of $\T$, we have
\begin{gather}
   \sum^{m}_{e=1}l_e=|\D|,\label{eq:5_20_6}
\end{gather}
By \eqref{eq:5_20_1}, we obtain $l_e \leq R'_e,\:  \forall e \in [m]$, % we have that
%\begin{gather}
%l_e \leq R'_e,\:  \forall e \in [m] ,\label{eq:5_20_7}
%\end{gather}
and thus,
\begin{gather}
  \sum^{m}_{e=1}l_e \leq \sum^{m}_{e=1}R'_e
  .\label{eq:5_20_8}
\end{gather}
In addition, from \eqref{4-22-2222}, we have
\begin{gather}
  \sum^{m}_{e=1}R'_e \leq |\A_{i_1}\cup\A_{i_2} \cup \hdots \cup \A_{i_m}|.
  \label{eq:5_20_9}
\end{gather}
Combining \eqref{eq:5_20_5}, \eqref{eq:5_20_6}, \eqref{eq:5_20_8} and \eqref{eq:5_20_9}, \eqref{eq:5_20_3} is true.
Therefore there exists a system of distinct representative. Consequently, the existence of the $N$-tuple is guaranteed. And then simply we designate
\begin{gather}
    \Z^{*}_i \triangleq \{x_{i,j} |  j \in [R'_i] \},\:\:\: \forall i \in [K] .\label{eq:5_20_4}
\end{gather}
Hence, the separate  assignment of $\Z^{(\pi_i)}_i$ sets is possible.

\section{Proof of Lemma 2}\label{A_3}
\begin{IEEEproof}
 Consider the following matrix
 \begin{align}
\mathbf{V}= \diag(\pmb{\zeta}^\intercal) \mathbf{C}+\mathbf{D},
\label{eq:6_63-100}
\end{align}
where $\mathbf{C}, \: \mathbf{D} \in \F^{n\times n}$, and
\begin{align}
        \pmb{\zeta}=
        \begin{bmatrix}  \zeta_1 & \zeta_2 & \hdots &\zeta_n \end{bmatrix}^\intercal
        &\in \F^{n\times1}.
\end{align}
We  define,
\begin{align}
        \I_i &\triangleq [i+1:n],\: \forall i\in [n],\\
    f(\S) &\triangleq [n]\backslash \S.
\end{align}
By contradiction, assume that there is no $\pmb{\zeta} $ vectors that results in $|\mathbf{V}|\neq0$, which means that $|\mathbf{V}|$ is identically equal to zero in terms of ${\zeta}_i,\: i \in [n]$s.
By co-factor expansion of the $|\mathbf{V}|$, we obtain,
\begin{align}
|\mathbf{V}|&= \sum^{n}_{r=1} (-1)^{r+1} [ {\zeta}_1c_{1,r}+
    d_{1,r}]|\mathbf{V}^{}({\I_{1},f(\{r\})}|\label{eq:6_65}
    \\ &={\zeta}_1 (\sum^{n}_{r=1} (-1)^{r+1}c_{1,r}|\mathbf{V}^{}(\I_{1},f(\{r\})|
    )+    \sum^{n}_{r=1}(-1)^{r+1} d_{1,r}|\mathbf{V}^{}({\I_{1},f(\{r\})})| =0.\label{eq:6_66}
\end{align}
Since $|\mathbf{V}|$ is to be  identically equal to zero and $d_{1,r},\: r \in [n]$ are independent from ${\zeta}_1$, we have
\begin{align}
\sum^{n}_{r=1} (-1)^{r+1}d_{1,r}|\mathbf{V}^{}({\I_{1},f(\{r\}))}|
    =0. \label{eq:6_67}
\end{align}
%which means that the multiplicative statement of ${\zeta}_1$ must be identically equal to zero.
Also from the co-factor expansion of the determinant of any $\mathbf{V}^{}({\I_i,\J}), i\in [n],\: \S\subseteq [n]$ we have:
\begin{align}
   |\mathbf{V}^{}({\I_i,\J})|= \sum^{n}_{s=1,s\notin \J} (-1)^{s+q}{\zeta}_{j+1}c_{j+1,s}|\mathbf{V}^{}({\I_{i+1},\J\cup\{s\}})|+
   \sum^{n}_{s=1,s\notin \J} (-1)^{s+q}d_{j+1,s}|\mathbf{V}^{}({\I_{i+1},\J\cup\{s\}})|,\label{eq:6_68}
\end{align}
where $q \in\{0,1\}$ is chosen according to the original expansion formula. In other words, the alternating criteria of the summation for every co-factor must be preserved. Substituting $i=r,j =1$ in \eqref{eq:6_68} and then combining with \eqref{eq:6_67}, we obtain
\begin{align}
&\sum^{n}_{r=1} (-1)^{r+1}c_{1,r}[ \sum^{n}_{s=1,s \neq r} (-1)^{s+q_1}{\zeta}_2c_{2,s}+
   \sum^{n}_{s=1,s \neq r}(-1)^{s+q_1} d_{2,s}]|\mathbf{V}^{}({\I_{r+1},f(\{r,s\})})|
    \label{eq:6_70-100}\\
   &= {\zeta}_2 \sum^{n}_{r=1} (-1)^{r+1}c_{1,r}[ \sum^{n}_{s=1,s \neq r} (-1)^{s+q_1}c_{2,s}|\mathbf{V}^{}({\I_{r+1},f(\{r,s\})})|\\&\quad+
   \sum^{n}_{s=1,s \neq r}(-1)^{s+q_1} d_{2,s}|\mathbf{V}^{}({\I_{r+1},f(\{r,s\})})|]\\&=0.\label{eq:6_69}
\end{align}
Since $u_{2,s},\: s \in \J$ are independent from ${\zeta}_2$, we  read
\begin{align}
  \sum^{n}_{r=1} (-1)^{r+1}c_{1,r}[ \sum^{n}_{s=1,s \neq r} (-1)^{s+q_1}c_{2,s}|\mathbf{V}^{}({\I_{r+1},f(\{r,s\})})|]=0.\label{eq:6_70}
\end{align}
Using \eqref{eq:6_68} and applying the similar substitution for another $n-2$ times, we conclude that
\begin{align}
    \sum^{n}_{r=1} (-1)^{r+1}c_{1,r}[ \sum^{n}_{s'=1,s \in \J'} (-1)^{s+q_1}c_{2,s}[\hdots|\mathbf{V}^{}({\I_{n-1},f(\S_{r'})})|]]=0,\label{eq:6_71}
\end{align}
where $\S_{r'}=f([n] \backslash \{r'\}),\: r' \in [n]$. From definition of ${V}({\I_{n-1},\S})$, we have
\begin{equation}
    {V}({\I_{n-1},\S_{r'}})=  {\zeta}_n c_{n,r'} + d_{n,r'}.\label{eq:6_72}
\end{equation}
Substituting \eqref{eq:6_72} in \eqref{eq:6_71} we get,
\begin{align}
   &\sum^{n}_{r=1} (-1)^{r+1}c_{1,r}[ \sum^{n}_{s'=1,s \in \J'} (-1)^{s+q_1}c_{2,s}[\hdots ({\zeta}_n c_{n,r'} + d_{n,r'})]]\label{eq:6_74}\\&=
   {\zeta}_n \sum^{n}_{r=1} (-1)^{r+1}c_{1,r}[ \sum^{n}_{s'=1,s \in \J'} (-1)^{s+q_1}c_{2,s}[\hdots  c_{n,r'}]]  \\&+\sum^{n}_{r=1} (-1)^{r+1}c_{1,r}[ \sum^{n}_{s'=1,s \in \J'} (-1)^{s+q_1}c_{2,s}[\hdots d_{n,r'}]] \\&=0.\label{eq:6_75}
\end{align}
Since $d_{n,r'},\: r' \in [n]$ are independent from ${\zeta}_{r'}$, we obtain
\begin{align}
  \sum^{n}_{r=1} (-1)^{r+1}c_{1,r}[ \sum^{n}_{s'=1,s \in \J'} (-1)^{s+d_1}c_{2,s}[\hdots  c_{n,r'}]]= 0.\label{eq:6_76}
\end{align}

Now, let
\begin{align}
\mathbf{C}=
\left[\begin{array}{cccc}
c_{1,1}
 &
 c_{1,2}  &
\hdots & c_{1,n}
\\
 c_{2,1}
 &
 c_{2,2}  &
\hdots & c_{2,n}
\\
{\vdots}
 &
{\vdots}&
{\vdots} & {\vdots}\\
c_{n,1}
 &
c_{n,2}  &
\hdots & c_{n,n}
\end{array}\right]\label{eq:6_80}
\end{align},
and expand its determinant as,
\begin{align}
  |\mathbf{C}|= \sum^{n}_{r=1} (-1)^{r+1}c_{1,r}[ \sum^{n}_{s'=1,s \in \J'} (-1)^{s+q_1}c_{2,s}[\hdots  c_{n,r'}]].\label{eq:6_82}
\end{align}
Noting \eqref{eq:6_82}, \eqref{eq:6_76} implies that the determinant of $\mathbf{C}$ is zero which is a contradiction.
Therefore, $|\mathbf{V}|$ is not identically equal to zero and there exists some ${\zeta}_i \in \mathbb{F}^{}_ ,\: i\in [n]$ that makes the $\mathbf{V}$ matrix non-singular.
 \end{IEEEproof}

\end{document}